\documentclass[a4paper,12pt]{article}
\usepackage{fullpage}
\usepackage{amsmath}
\usepackage{amssymb}
\usepackage{natbib}

\usepackage{authblk}
\title{Structural Anisotropy Stabilises Asymmetric Beating in Instability Driven Filaments}
\author[,1]{Bethany Clarke\thanks{b.clarke21@imperial.ac.uk}}
\author[,2]{Yongyun Hwang\thanks{y.hwang@imperial.ac.uk}}
\author[,1]{Eric E Keaveny\thanks{e.keaveny@imperial.ac.uk}}

\affil[1]{Department of Mathematics, Imperial College London, SW7 2AZ, UK}
\affil[2]{Department of Aeronautics, Imperial College London, SW7 2AZ, UK}

\usepackage{setspace} 
\usepackage{graphicx}% Include figure files
\usepackage{dcolumn}% Align table columns on decimal point
\usepackage{bm}% bold math
\usepackage{caption}
\usepackage{subcaption}
\usepackage{comment}
\usepackage{xcolor}
\usepackage{cite}
\usepackage{array}
\usepackage{multirow}
\usepackage{setspace}
\begin{document}

\date{\today}

\maketitle

\begin{abstract}
Asymmetries and anisotropies are widespread in biological systems, including in the structure and dynamics of cilia and eukaryotic flagella. These microscopic, hair-like appendages exhibit asymmetric beating patterns that break time-reversal symmetry needed to facilitate fluid transport at the cellular level.  The intrinsic anisotropies in ciliary structure can promote preferential beating directions, further influencing their dynamics.  In this study, we employ numerical simulation and bifurcation analysis of a mathematical model of a filament driven by a follower force at its tip to explore how intrinsic curvature and direction-dependent bending stiffnesses impact filament dynamics.  Our results show that while intrinsic curvature is indeed able to induce asymmetric beating patterns when filament motion is restricted to a plane, this beating is unstable to out of plane perturbations. Furthermore, we find that a 3D whirling state seen for isotropic filament dynamics can be suppressed when sufficient asymmetry or anisotropy are introduced. Finally, for bending stiffness ratios as low as 2, we demonstrate that combining structural anisotropy with intrinsic curvature can stabilise asymmetric beating patterns, highlighting the crucial role of anisotropy in ciliary dynamics.
\end{abstract}

%\section{Introduction}
%In the isotropic follower force case, all dynamics are planar. In particular, the initial buckling bifurcation is a double Hopf bifurcation; a codim 2 bifurcation, in which one of the parameters is the nondimensional follower force strength, $f$. There must therefore be a second parameter, which controls the asymmetry of the system - breaking this symmetry would lead to two separate Hopf bifurcations. In this work, we investigate two such potential parameters; $\alpha,$ which controls the preferred curvature of the filament, and $\beta$, which is defined to measure the bending modulus anisotropy. The previous isotropic case considers $\alpha=0, \beta=1$. In the first two sections we see how the state space varies while we hold the other variable fixed, before finally considering how the state space changes as we vary both.

%\input{tex_files/preferred_curvature}
%\input{tex_files/bending_modulus}
%\input{tex_files/curvature_and_modulus}

\section{Introduction}
Asymmetry is ubiquitous in nature across multiple scales, from the chirality of molecules \citep{Garbacz2024IntroductionChirality} to the left-right asymmetry in the placement of internal organs in the human body \citep{Hirokawa2009Left-rightFlow.}. Many animals utilise asymmetry for mating or survival, such as flatfish like flounders and halibut, which have both eyes on one side of the head to allow them to lie flat against the ocean floor and avoid predators \citep{Palmer2009AnimalAsymmetry}. Anisotropy, here referring to the direction dependence of mechanical response of a material, is also common in nature, a simple example being the human hand, which is easier to bend in one direction than in any other direction. Fibrous materials such as wood are anisotropic due to the natural grain formed by fibre alignment, meaning the strength and flexibility of these materials are direction-dependent \citep{Kretschmann2010ChapterWood}. 
 
Cilia and eukaryotic flagella are hairlike appendages which exhibit both asymmetric and anisotropic properties. These filaments are responsible for driving fluid flow at the microscale \citep{Gibbons1981CiliaEukaryotes}. They can be found on the surfaces of swimming cells and microorganisms such as \textit{Volvox} \citep{Pedley2016SquirmersSwimming}, collectively beating to allow the algae to navigate through flows \citep{Suarez2006SpermTract, Bennett2015AChlamydomonas}, or lining epithelial cells in our own bodies to facilitate fluid transport, such as mucus in the airways \citep{Smith2008ModellingClearance}, or cerebrospinal fluid in the brain \citep{Worthington1966CILIARYSURFACES,Guldbrandsen2014In-depthCSF-PR.,Zhang2015AProteome}. Symmetry breaking is essential for fluid transport at this scale \citep{Purcell1977LifeNumber}. Accordingly, cilia often undergo asymmetric beating oscillations, to ensure net fluid motion and cellular transport. Respiratory cilia for example, beat with distinct fast, low-curvature effective strokes, and high-curvature, slower, recovery strokes \citep{Chilvers2000AnalysisMethods,Jorissen2023CiliaTransport}. Nodal cilia in the embryo, on the other hand, undergo tilted whirling motions \citep{Nonaka1998RandomizationProtein,Smith2008FluidCilia,Smith2011MathematicalCilia}, that direct fluid flow to the left and hence breaking left-right symmetry in the embryo. 

Such asymmetries in ciliary dynamics are thought to be the result of the internal structure of cilia known as the axoneme. While the configuration of the axoneme can vary \citep{Chaaban2017APolymers}, motile cilia most commonly have a 9+2 axonemal structure \citep{Gibbons1981CiliaEukaryotes,Marshall2006Cilia:Antenna,Gilpin2020TheFlagella}, consisting of a central pair of microtubules, surrounded by 9 pairs of microtubules, called microtubule doublets. The microtubule doublets are held relative to each other by structural proteins called nexin. The bridge between one pair of doublets is known to be much stiffer than the others \citep{Gibbons1981CiliaEukaryotes,AFZELIUS1959ElectronFixative.}, indicating that this pair have limited motion relative to each other. The orientation of the central pair and the stiffer bridge suggest a plane in which it is easier for the axoneme to bend. It has been observed that the beat plane for cilia aligns with this plane \citep{Gibbons1981CiliaEukaryotes}, suggesting that the direction of the cilium beat could be a result of anisotropy in axonemal structure. In previous numerical studies modelling cilia, this has been included explicitly by considering a stiffer bridge between this pair of doublets \citep{Han2018SpontaneousCilia}, or implicitly by increasing the perpendicular bending stiffness \citep{Rallabandi2022Self-sustainedFlagellum}. It has also often been used as motivation for restricting dynamics to be entirely planar \citep{Bayly2016SteadyFlagella}.

While these modelling choices can reproduce planar beating, the resulting dynamics are often symmetric and do not reproduce differences in the effective and recovery strokes observed experimentally. Additional asymmetry must be considered to generate more accurate simulations which match experiments. Some works achieve this by incorporating constraints on material properties of the filament, such as curvature-dependent bending stiffness \citep{Han2018SpontaneousCilia} in which the bending stiffness can take different values depending on the local curvature of the filament's centerline, or through biased kinetics of dynein motors \citep{Chakrabarti2019SpontaneousMicrofilaments,Chakrabarti2019HydrodynamicFilaments,Chakrabarti2022ACilia}. A widely accepted approach, however, includes introducing an intrinsic curvature to the filament \citep{Chakrabarti2019SpontaneousMicrofilaments,Chakrabarti2019HydrodynamicFilaments,Sartori2016DynamicFlagella,Cass2023TheFlagella,Wang2023Generalized-NewtonianFilament}, in which it is assumed that the rest state of the cilium has finite curvature. These numerical studies are successful in demonstrating realistic cilia beats, with a close resemblance to the beats observed in nature, but are often restricted to two dimensional dynamics. Further studies are necessary to determine whether these dynamics persist when the filament is free to deform in any direction.

In this paper we explore the effect of asymmetries and anisotropies in a 3D active filament model. To drive filament oscillations, we utilise the widely-studied follower force model \citep{Bayly2016SteadyFlagella,DeCanio2017,Ling2018Instability-drivenMicrofilaments,Fily2020BucklingFilaments,Fatehiboroujeni2018NonlinearDrag,Fatehiboroujeni2021Three-dimensionalFlipping,Clarke2024BifurcationsFilaments,Schnitzer2025OnsetFilament}, whereby a single compressive force is applied to the tip of the filament, against the tangent. While the follower force model does not represent the sliding-mechanism known to generate ciliary beating, this model has been shown to exhibit beating and whirling dynamics for sufficient forcing values, which are reminiscent of cilia beating patterns observed in nature. As such, the follower force model has been used to explore instabilities underpinning cilia dynamics \citep{Bayly2016SteadyFlagella,Woodhams2022GenerationModels}, understand the influence of fluid rheology on cilia's fluid-pumping ability \citep{Wang2023Generalized-NewtonianFilament,Link2024EffectFilament}, study filament beat coordination and collective dynamics \citep{Westwood2021CoordinatedSurfaces}, and model biofilaments in cytoplasmic streaming in \textit{Drosophila} oocytes \citep{Stein2021SwirlingCytoskeleton} and actin-myosin dynamics \citep{Sekimoto1995SymmetrySystem}. In the next section, we describe the active filament model we utilise to generate simulations, and the computational tools we employ to find steady and periodic states, and to analyse their stability. Next, we include asymmetry through a non-trivial intrinsic curvature to an otherwise-isotropic filament, and establish the dynamics in both 2D, matching those observed in the literature, and 3D. We find that the asymmetric planar beating found in 2D is inherently unstable in the full 3D model, and that the whirling solution can be eliminated entirely if the imposed asymmetry is large enough. Following this, we discuss the effect of direction-dependent bending stiffness in the absence of intrinsic curvature, showing how whirling can again be eradicated if the anisotropy in the bending stiffnesses is sufficiently large. Finally, we investigate how this anisotropy can be exploited on filaments with intrinsic curvature to recover and stabilise asymmetric beating dynamics in 3D.  In particular, we show that bending stiffness ratios as low as 2 are sufficient to stabilise asymmetric beating, a value agreeing with structural anisotropies that have been suggested for cilia \citep{Ishijima1994FlexuralFlagella,Lindemann2016FunctionalFlagellum,Rallabandi2022Self-sustainedFlagellum}.

\section{Numerical methods}
%\begin{itemize}
%    \item Filament model (including how to incorporate intrinsic curvature/different bending stiffnesses, define $\kappa_0, \beta$)
%    \item JFNK
%    \item Stability analysis
%\end{itemize}

\begin{figure}
    \centering
    \includegraphics[width=0.8\linewidth]{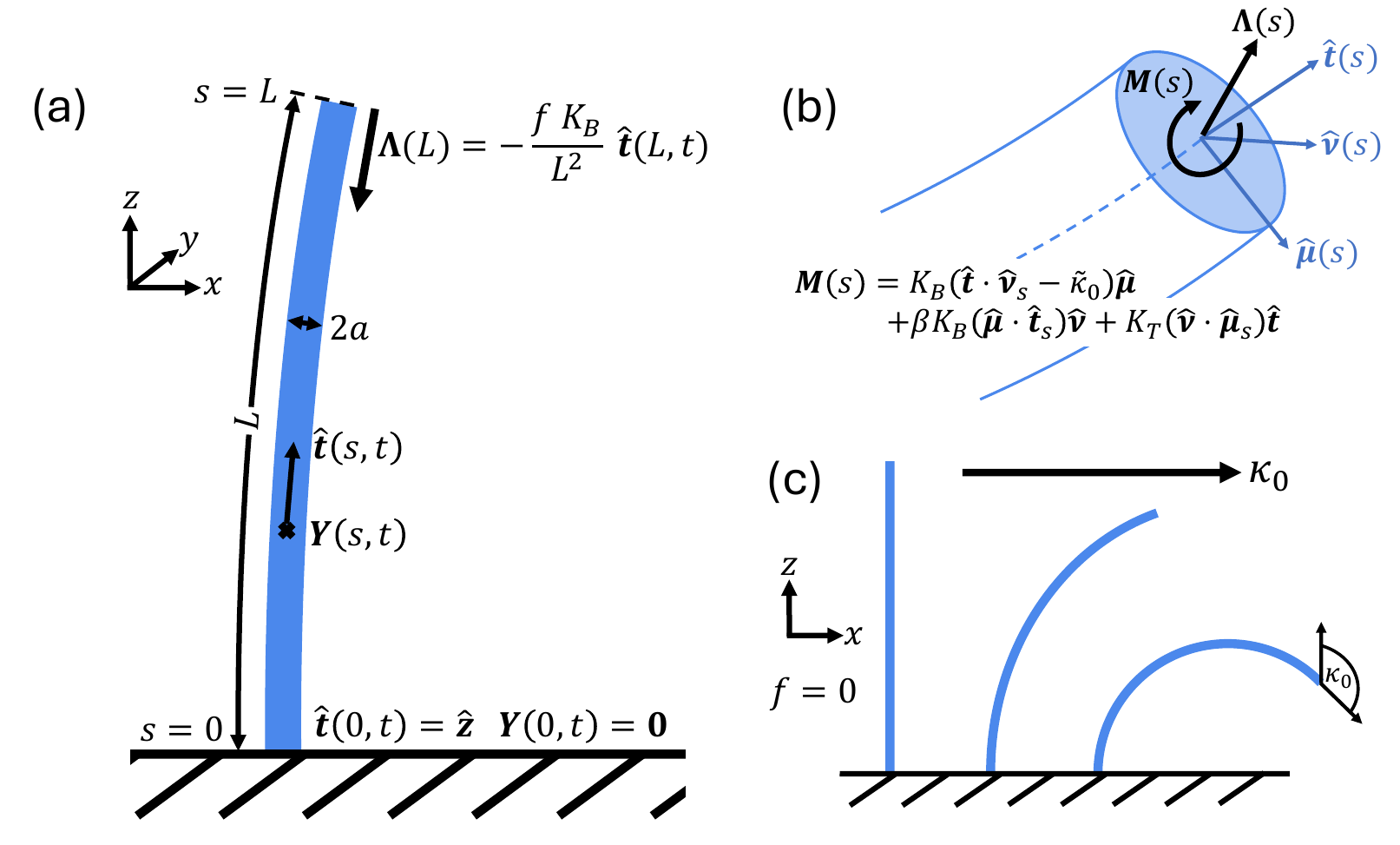}
    \caption{Schematics of the model. (a) We consider a filament of length $L$ and radius $a$, clamped at its base $(s=0)$ to an infinite no-slip planar wall, and driven by a follower force imposed at the free end $(s=L).$ (b) The bending stiffness anisotropy and intrinsic curvature are defined through the bending moment, which act on the filament cross-section. It is expressed in terms of the filament's material frame, $(\hat{\bm{t}},\hat{\bm{\mu}},\hat{\bm{\nu}})$. We emphasise that the filament cross-section is circular. (c) The intrinsic curvature, $\kappa_0 = L\tilde{\kappa_0}$ is defined to be the rest-curvature of the filament under zero forcing. }
    \label{fig:method}
\end{figure}
The methods employed here largely follow those described in \citep{Scholler2021MethodsFilaments,Clarke2024BifurcationsFilaments}, and we provide a brief description below for completeness.

\subsection{Filament Model}\label{sec:filament_model}
The filament has length $L$ and cross-sectional radius $a$, and is clamped at its base to a no-slip planar surface.  The filament is forced at its free end by a follower force, a compressive tangential force, as shown in Figure \ref{fig:method}a. The filament centerline, $\bm{Y}(s,t),$ is parameterised by arc-length, $s\in [0,L]$, and evolves over time, $t\in [0,\infty)$. Filament deformation is described by the orthonormal local basis $\{\bm{\hat{t}}(s,t),\bm{\hat{\mu}}(s,t),\bm{\hat{\nu}}(s,t)\}$. For the filament dynamics, we use the framework outlined in \citep{Scholler2021MethodsFilaments}. Due to the small length scales and velocity scales associated to ciliary motion, the Reynolds number is small and so we neglect the inertia of the fluid and the filament. Consequently, force and moment balances along the filament length yield
\begin{align}
     \frac{\partial \bm{\Lambda}}{\partial s}+\bm{f}^H &=\bm{0}, \label{eq:continuous_equations1}\\
     \frac{\partial \bm{M}}{\partial s}+\bm{\bm{\hat{t}}} \times \bm{\Lambda}+\bm{\tau}^H &=\bm{0}, \label{eq:continuous_equations2}
\end{align}
where $\bm{f}^H(s,t)$ and $\bm{\tau}^H(s,t)$ are the forces and torques per unit length, respectively, exerted by the surrounding fluid, and  $\bm{\Lambda}(s,t)$ and $\bm{M}(s,t)$ are the internal force and moment, respectively, acting on the filament cross-section (see Figure \ref{fig:method}b). The internal forces act as Lagrange multipliers enforcing the kinematic constraint,
\begin{equation}
     \frac{\partial \bm{Y}}{\partial s} =\bm{\hat{t}}, \label{eq:continuous_equations3}    
\end{equation}
and the internal moments are given by the constitutive relation \citep{Scholler2021MethodsFilaments},
\begin{equation}\label{eq:moment_continuous}
    \bm{M}(s,t) = K_B \left(\hat{\bm{t}} \cdot \frac{\partial \hat{\bm{\nu}}}{\partial s} - \tilde{\kappa}_0\right) \hat{\bm{\mu}} + \beta K_B\left(\hat{\bm{\mu}} \cdot \frac{\partial \hat{\bm{t}}}{\partial s} \right) \hat{\bm{\nu}}+ K_T \left( \hat{\bm{\nu}} \cdot \frac{\partial \hat{\bm{\mu}}}{\partial s}\right) \hat{\bm{t}},
\end{equation}
where $\tilde{\kappa}_0 = \kappa_0/L$ defines the intrinsic curvature, such that the rest configuration of the filament is a planar arc with constant curvature $\kappa_0,$ as sketched in Figure \ref{fig:method}c. Therefore, larger values of $\kappa_0$ correspond to a higher curvature rest configurations for the filament. Also defined in Eq. \eqref{eq:moment_continuous} are $K_B$, the filament bending stiffness in the direction of intrinsic curvature, $\beta K_B$, the bending stiffness in the perpendicular plane, and $K_T$, the twisting stiffness. Choosing $\beta>1$ leads to preferred bending about the $\hat{\bm{\mu}}-$axis, and $\beta<1$ leads to preferred bending about the $\hat{\bm{\nu}}-$axis. Correspondingly, setting $\beta \rightarrow \infty$ should restrict the filament dynamics to the $(\hat{\bm{\nu}},\hat{\bm{t}})$-plane entirely. Choosing $\beta = 1, \kappa_0 = 0$ corresponds to the isotropic, symmetric case considered in \citep{Clarke2024BifurcationsFilaments}.

The base of the filament $(s=0)$ is clamped to a no-slip planar surface, fixing the position and local orientation of the base to be $\bm{Y}(0,t) = \bm{0}$ and $(\hat{\bm{t}}(0,t),\hat{\bm{\mu}}(0,t),\hat{\bm{\nu}}(0,t)) = (\hat{\bm{z}},\hat{\bm{y}},\hat{\bm{x}})$. The distal end $(s=L)$ is moment-free, $\bm{M}(L,t)=\bm{0}$, and is driven by the follower force,
\begin{equation}\label{eq:followerforce_BC}
    \bm{\Lambda}(L,t) = -\frac{f K_B}{L^2} \hat{\bm{t}}(L,t),
\end{equation}
where $f$ is the non-dimensional follower force strength. Through this non-dimensionalisation, the length is incorporated in the non-dimensional parameter, $f$. The impact of filament aspect ratio, $L/a$, in the symmetric and isotropic case ($\kappa_0 = 0$ and $\beta = 1$) was investigated in \citep{Clarke2024BifurcationsFilaments}. We note that an alternative definition of the non-dimensional follower force can be reached by balancing with the elastic force in the $\hat{\bm{\nu}}$-direction. This would lead to a rescaled non-dimensional follower force, $\tilde{f} = \beta f$.

We discretise the filament into $N$ segments of length $\Delta L$, such that segment $i$ has centre $\bm{Y}_i$ and an orientation defined by the local frame, $\{\bm{\hat{t}}_i,\bm{\hat{\mu}}_i,\bm{\hat{\nu}}_i\}$ for $i=1,...,N$. We discretise Eqs \eqref{eq:continuous_equations1}, \eqref{eq:continuous_equations2} and \eqref{eq:continuous_equations3}, using central differencing and rearrange to obtain
\begin{align}
    \bm{F}_i^{C}+\bm{F}_i^H &=\bm{0},\label{eq:forcebal_discrete} \\
    \bm{T}_i^{E}+\bm{T}_i^{C}+\bm{T}_i^H &=\bm{0} \label{eq:torquebal_discrete},
\end{align}
for segment $i=1,...,N$, and
\begin{equation}    
    \bm{Y}_{i+1}-\bm{Y}_i-\frac{\Delta L}{2}\left(\bm{\hat{t}}_i+\bm{\hat{t}}_{i+1}\right) =\bm{0},\label{eq:inextensibility_discrete}
\end{equation}
for $i=1,...,N-1$. Here $\bm{T}_i^{E} = \bm{M}_{i+1/2} - \bm{M}_{i-1/2}$ is the elastic torque, $\bm{F}_i^{C} = \bm{\Lambda}_{i+1/2} - \bm{\Lambda}_{i-1/2}$ and $\bm{T}_i^C =(\Delta L/2) \hat{\bm{t}}_i \times (\bm{\Lambda}_{i+1/2} + \bm{\Lambda}_{i-1/2}) $ are the constraint forces and torques respectively, and $\bm{F}_i^H = \Delta L \bm{f}_i^H$ and $\bm{T}_i^H = \Delta L \bm{\tau}_i^H$ are the hydrodynamic force and torque respectively. The internal forces, $\bm{\Lambda}_{i+1/2},$ act as Lagrange multipliers enforcing the constraint, \eqref{eq:inextensibility_discrete}. The internal moments, $\bm{M}_{i+1/2}$, are defined by a discretised version of the constitutive law given in Eq \eqref{eq:moment_continuous}.

As we neglect fluid inertia due to the low Reynolds number, the fluid velocity is governed by the Stokes equations. Therefore, the translational and rotational velocities of the segments are linearly related to the hydrodynamic forces and torques on the segments through the mobility matrix, $\mathcal{M}$. Explicitly, this can be expressed via the mobility relation:
\begin{equation}\label{eq:mobility_matrix_OG}
    \left(\begin{array}{l} \bm{V} \\ \bm{\Omega}\end{array}\right)= \mathcal{M} \left(\begin{array}{l}-\bm{F}^{H} \\ -\bm{T}^{H}\end{array}\right),
\end{equation}
where $\mathcal{M}$ is the $6N \times 6N$ mobility matrix, $\bm{V}$ and $\bm{\Omega}$ are $6N \times 1$ vectors containing the translational and angular velocities of all segments respectively, and $\bm{F}^H$, $\bm{T}^H$ are $6N\times 1$ vectors containing the hydrodynamic forces and torques of all segments, respectively, which are defined through the force and moment balances, Eqs \eqref{eq:forcebal_discrete} and \eqref{eq:torquebal_discrete}. In this work, we use the Rotne-Prager-Yamakawa (RPY) tensor \citep{Wajnryb2013GeneralizationTensors}, adapted to include the no-slip condition at $z=0$ \citep{Swan2007SimulationBoundary}, to define the mobility matrix, $\mathcal{M}$. Here, each segment is modelled as a particle with hydrodynamic radius equal to the cross-sectional radius of the filament, $a$. Various hydrodynamic models have been used in this context. The follower force model appears to give rise to the same dynamics in the forcing ranges we consider, regardless of the details of the hydrodynamic model \citep{Ling2018Instability-drivenMicrofilaments}. For a comparison between RPY and nonlocal slender body theory, we direct the reader to \citep{Maxian2021Integral-basedFlow}.

After solving for the translational and rotational velocities of each segment, we integrate the system forwards in time. To understand the evolution of segment orientations, we utilise unit quaternions, $\bm{q} = (q_0,q_1,q_2,q_3)$ with $||\bm{q}||=1$ \citep{Scholler2021MethodsFilaments,Hanson2006VisualizingQuaternions}. Segment $i$ is defined to have a corresponding unit quaternion, $\bm{q}_i$, which maps the standard basis, $(\hat{\bm{x}},\hat{\bm{y}},\hat{\bm{z}})$, to the segment's local basis at time $t,$ $(\hat{\bm{t}}_i(t)\hat{\bm{\mu}}_i(t)\hat{\bm{\nu}}_i(t))$, through a rotation matrix, $\bm{R}(\bm{q})$ (see \citep{Scholler2021MethodsFilaments} for further details). The evolution of each quaternion is described by
\begin{equation}\label{eq:discrete_eq_2}
    \frac{d\bm{q}_i}{d t} = \frac{1}{2}\left( 0, \bm{\Omega}_i \right) \bullet \bm{q}_i,
\end{equation}
where $\bullet$ is the quaternion product, defined for two quaternions $\bm{p} = (p_0, \tilde{\bm{p}})$ and $\bm{q} = (q_0, \tilde{\bm{q}})$ as $\bm{p} \bullet \bm{q} = (p_0 q_0 - \tilde{\bm{p}}\cdot \tilde{\bm{q}}, \; p_0\tilde{\bm{q}} + q_0 \tilde{\bm{p}} + \tilde{\bm{p}} \times \tilde{\bm{q}})$.

Pairing Eq \eqref{eq:discrete_eq_2} with an equation describing the evolution of segment positions, namely
\begin{equation}\label{eq:discrete_eq_1}
    \frac{d\bm{Y}_i}{d t} = \bm{V}_i,
\end{equation}
we can discretise our system temporally using a second-order, geometric backward differential formula scheme, as described in \citep{Scholler2021MethodsFilaments}. These discretised equations, as well as the constraint given by \eqref{eq:inextensibility_discrete}, define a nonlinear system which we can solve iteratively using Broyden's method \citep{Broyden1965AEquations}.  

In this work, the filament is discretised into $N=20$ segments of length $\Delta L = 2.2a$, such that the filament has length $L = N\Delta L = 44a.$ This separation has been chosen to avoid overlap between neighbouring segments when the filament bends. The twist stiffness is taken to be equal to the bending stiffness in the $\hat{\bm{\mu}}-$direction, $K_B=K_T$. In the isotropic case, where $\beta=1$ and $\kappa_0=0$, the magnitude of the twist has little effect on the dynamics \citep{Clarke2024BifurcationsFilaments}. We investigate the effect of varying three non-dimensional parameters;  the intrinsic curvature, $\kappa_0$, as in \eqref{eq:moment_continuous}; the bending anisotropy, quantified by $\beta,$ as in \eqref{eq:moment_continuous}; and the nondimensional follower force, $f,$ as defined in \eqref{eq:followerforce_BC}. We focus on ranges of $\kappa_0$ and $\beta$ that follow values explored in other studies \citep{Rallabandi2022Self-sustainedFlagellum,Chakrabarti2019SpontaneousMicrofilaments,Chakrabarti2019HydrodynamicFilaments,Chakrabarti2022ACilia,Sartori2016DynamicFlagella,Cass2023TheFlagella,Wang2023Generalized-NewtonianFilament,Lindemann2016FunctionalFlagellum}, and a range of $f$ which allows us to access and explore the whirling and beating states. All initial value problems (IVPs) use the initial condition of a vertically upright filament, with random perturbations to the force on each segment. All timescales are non-dimensionalised by the filament's relaxation time, $\tau$, the characteristic timescale associated with a bent filament returning to its undeformed configuration, which we obtain numerically as described in \citep{Clarke2024BifurcationsFilaments}. 
\subsection{Bifurcation and Stability Analyses}\label{Section:bifurcation_theory}
The filament model offers a method to generate numerical solutions given an initial configuration. In order to understand the bifurcations between different states that arise from the model, we need to first obtain these states and then assess their stability. We use a Jacobian Free Newton-Kyrlov subspace method (JFNK) to find the stable or unstable steady-state and time-periodic solutions, as described in \citep{Viswanath2007RecurrentTurbulence,Willis2019EquilibriaThem, Clarke2024BifurcationsFilaments}. JFNK seeks solutions which minimise the error between the initial state and the state after being advanced for a period of time. Therefore, we can only use this to identify steady and time periodic solutions - not, for instance, quasiperiodic or chaotic dynamics. The system is solved using Newton's method, where the Generalised Residual Method (GMRES) is employed in each Newton iteration to solve for the update. Once we have a solution for a single set of parameters, we can then use continuation to track the solution branch for different values of $f, \kappa_0$ or $\beta$, including values of these parameters for which the solution may be unstable. 

After finding the steady-state or time-periodic base state, we can analyse its stability numerically. This is achieved by using the Arnoldi method to establish the eigenvalues of the system's Jacobian, for steady state solutions, or the Floquet matrix exponent, if the state is time-periodic. 
Further details of this method can be found in \citep{Clarke2024BifurcationsFilaments}. 

It may be that two solutions are stable for the same values of parameters. If this is the case, we expect for there to be an unstable state which lies on the boundary of the two stable solutions in the state space. We can identify the unstable solution by using the bisection algorithm outlined in \citep{Skufca2006EdgeFlow}, which iteratively bisects between initial conditions leading to either stable state. While this method allows us to identify an unstable solution lying on the boundary surface (separatrix) between the stable states in the state space, we only observe this transiently, before the filament converges to one of the stable states. Furthermore, if we expect multiple unstable solutions to exist for a set of parameters, bisection can only compute a solution stable (or attractive) along the directions tangential to the boundary surface between the two stable states.

\section{Intrinsic Curvature in isotropic filaments}\label{sec:curvature}
%\begin{itemize}
%    \item In 2D find beating through a Hopf bifurcation, same value regardless of $\kappa_0,$ but with asymmetry like \citep{Wang2023Generalized-NewtonianFilament} close to the bifurcation. This is unstable in 3D
%    \item In 3D: Lose double hopf bifurcation (steady state now changes with $f$). Out-of-plane mode becomes unstable earlier as we increase $\kappa_0$, in-plane mode becomes unstable at same value.
%    \item Observe P1 next (PB becomes P1, describe P1)
%    \item For smaller $\kappa_0,$ then observe whirling (asymmetry closer to initial bifurcation). Floquet reveals bistability between whirling and P1, implying a quasiperiodic branch existing between them. Whirling becomes unstable for P1
%    \item For larger $\kappa_0$, we lose whirling (describe bifurcation, link to distance between in-plane and out-of-plane modes becoming unstable in linear model)
%    \item Higher order modes become visible in top right of phase diagram (e.g. P2)
%\end{itemize}

\subsection{Planar Dynamics}
\begin{figure}[t]
    \centering
    \includegraphics[width=0.75\linewidth]{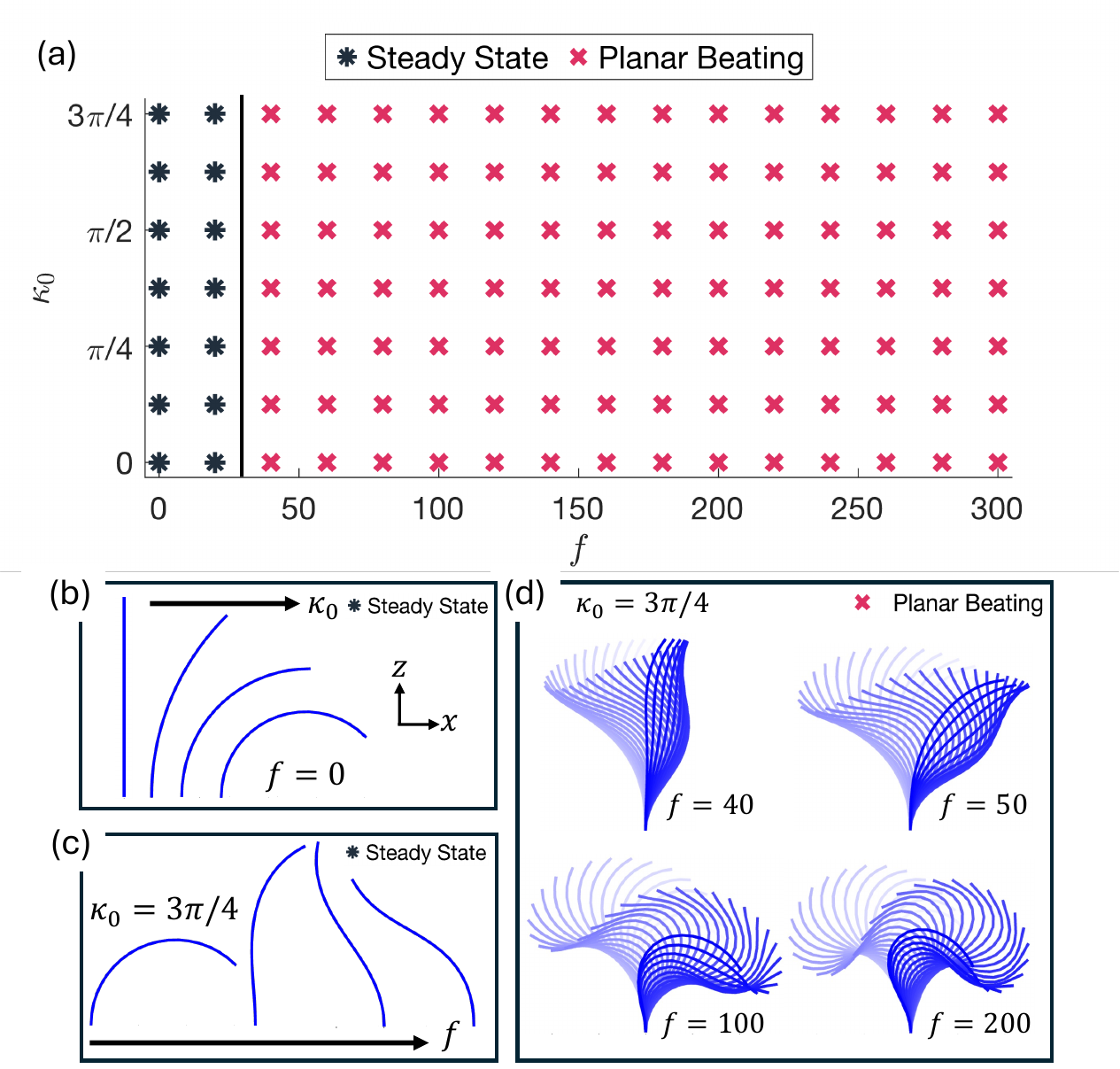}
    \caption{(a) Results from 2D numerical simulations of a filament with intrinsic curvature $(\kappa_0 \geq 0,$ $ \beta = 1)$. The black line is drawn to indicate the approximate boundaries of the solution regions. For $f<f^* \approx 35.3$, filaments remain in their steady non-trivial state which varies with (b) $\kappa_0$ (shown for $f=0, $ $\kappa_0 \in \{0,\pi/4,\pi/2,\pi/4 \}$) and (c) $f$ (shown for $\kappa_0 = 3\pi/4,$ $ f \in \{0,5,10,20\}$). For $f>f^*$ filaments undergo asymmetric oscillations for $\kappa_0>0$, as shown for (d) $\kappa_0=3\pi/4$.}
    \label{fig:curvature_phasediagram}
\end{figure}
We begin by considering an isotropic filament with finite intrinsic curvature, i.e. $\beta = 1,$ and $ \kappa_0 \neq 0$, with its motion restricted to the plane in which it is deformed by its intrinsic curvature. The corresponding phase diagram of the planar filament states in $\kappa_0-f$ space, where the plane is aligned with the filament's intrinsic curvature, is obtained by running one initial value problem for each set of parameters. Each initial value problem has an initial condition of a vertically upright filament with random, small perturbations to the forcing on a subset of the segments. We summarise the results in Figure \ref{fig:curvature_phasediagram}a. For this study we consider $\kappa_0 \in [0,3\pi/4]$. This range of values is appropriate as it includes values used in the literature in numerical studies \citep{Wang2023Generalized-NewtonianFilament}, in particular to recover dynamics similar to those observed experimentally \citep{Chakrabarti2019HydrodynamicFilaments,Sartori2016DynamicFlagella}. These IVPs reveal that we observe two types of behaviour; steady states below a critical value of the forcing, or self-sustained beating oscillations when this critical value is exceeded, matching observations in \citep{Wang2023Generalized-NewtonianFilament}.

For $\kappa_0=0$, the steady state is a vertically upright filament. However increasing $\kappa_0 \neq 0,$ the steady state becomes non-trivial. For $f=0,$ it is given by the prescribed intrinsic curvature, such that 
\begin{equation}
    \bm{t}(s) = (\sin(\kappa_0 s/L), 0, \cos(\kappa_0 s/L)), \;\;\; s\in[0,L],
\end{equation}
as shown in Figure \ref{fig:curvature_phasediagram}b.
When $f>0,$ this steady state changes; the tip of the filament is pushed by the follower force, until a new steady state is reached in which the elastic forces balance the follower force. This gives a non-trivial steady state, as shown for $\kappa_0=3\pi/4$ in Figure \ref{fig:curvature_phasediagram}c. 

When the force exceeds a critical value, the filament buckles and we observe planar oscillations. To find this critical value for each $\kappa_0,$ we track the steady state using JFNK and use stability analysis. For $\kappa_0=0,$ this reveals that the buckling occurs at $f = f^* \approx 35.3,$ through a Hopf bifurcation. For $\kappa_0>0$, the bifurcation occurs through the same mechanism, but we observe a small increase in the critical forcing of approximately $0.5\%$ from $\kappa_0=0$ to $\kappa_0=3\pi/4.$  For $\kappa_0 = 0,$ numerical simulations after the bifurcation show that the planar beating oscillations are symmetric. However, this symmetry is broken by intrinsic curvature; for $\kappa_0 > 0,$ simulations reveal asymmetric beating patterns as shown in Figure \ref{fig:curvature_phasediagram}d, and observed in \citep{Wang2023Generalized-NewtonianFilament}. We plot the frequencies of these solutions for various $\kappa_0$ in Appendix \ref{appen:frequencies}. Close to the bifurcation this asymmetry is strongest, but further away from the bifurcation, where the follower force is strong, the effect of intrinsic curvature is less pronounced and the beating becomes more symmetric.

\subsection{Non-planar Dynamics - Initial Value Problems}
\begin{figure}[t]
    \centering
    \includegraphics[width=0.75\linewidth]{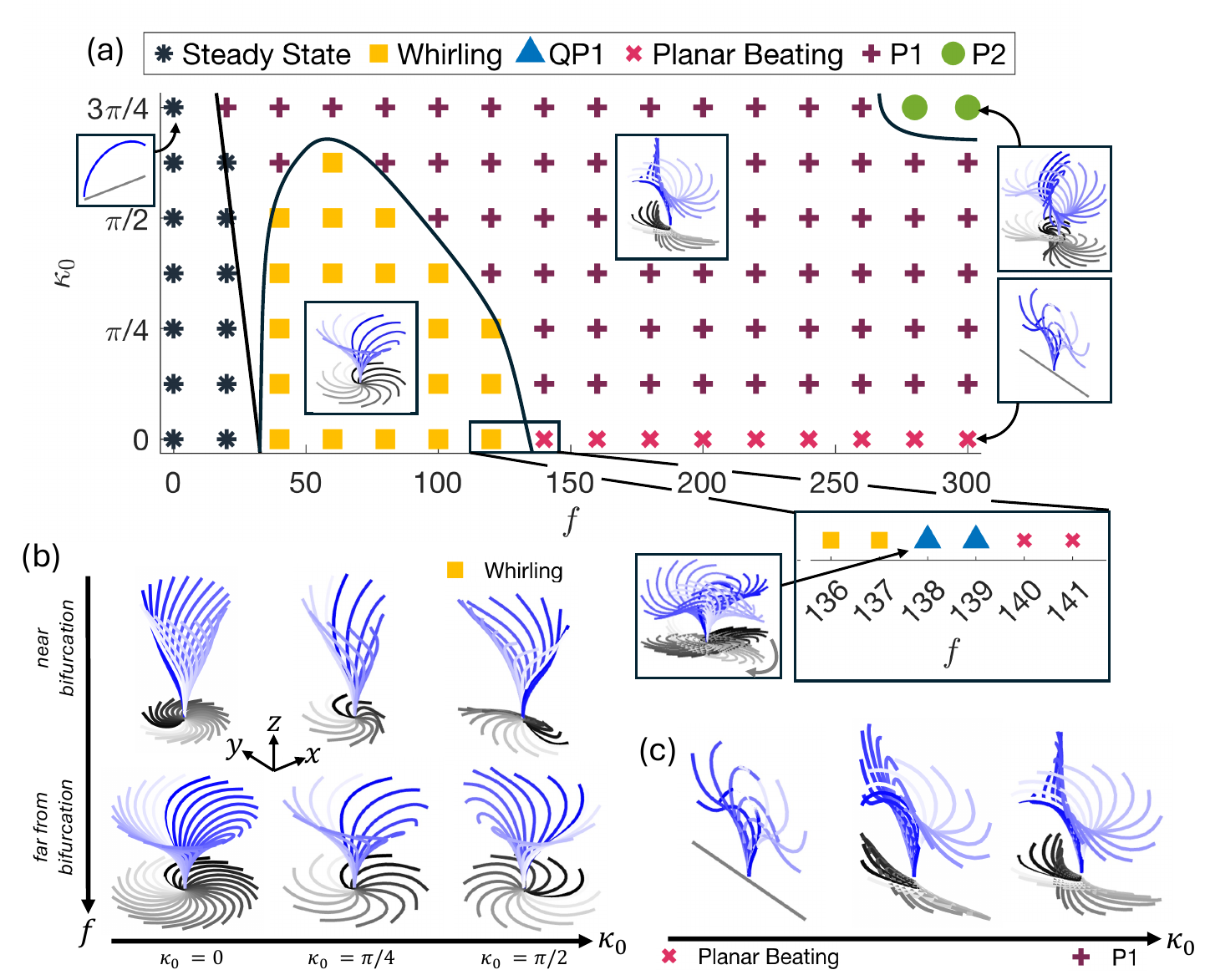}
    \caption{(a) Results of numerical simulations of a filament with intrinsic curvature in 3D $(\kappa_0 \geq 0, \beta = 1)$. The black lines are drawn to indicate the approximate boundaries of the solution regions. (b) Snapshots of filaments over one period undergoing whirling dynamics for three values of intrinsic curvature close to ($f=40$) and further from ($f=100$) the bifurcation. We see asymmetries in whirling for $\kappa_0>0$ are more pronounced close to the bifurcation, and for higher $\kappa_0.$ (c) Snapshots of filaments over one period for $f=200,$ increasing $\kappa_0$ from $0$ to $3\pi/4$. The dynamics change from planar beating for $\kappa_0 = 0$ (left) to P1 for $\kappa_0 > 0$ (right).}
    \label{fig:curvature_phasediagram_3D}
\end{figure}
When the filament is free to deform in any direction, we observe dramatic changes in filament dynamics as we vary $\kappa_0$ and $f$. We summarise the filament states in Figure \ref{fig:curvature_phasediagram_3D}a. For the smallest values of the forcing, we observe the same steady states as in the planar case. However, we now find that the filament exhibits fully 3D dynamics after buckling and the critical value of the forcing has a dependence on $\kappa_0$.

For $\kappa_0 = 0$ \citep{Ling2018Instability-drivenMicrofilaments,Clarke2024BifurcationsFilaments}, we observe the symmetric whirling state immediately after the buckling; a rigid body rotation of the centerline, with the filament tip tracing a circle over one period. Increasing the forcing further, whirling is replaced by a quasiperiodic solution which we term QP1 - an elliptical whirling which rotates unidirectionally. QP1 only exists for a small region of $f,$ before being replaced with planar beating. Planar beating is also observed in the 2D simulations, but is now symmetric (as $\kappa_0=0$) and can occur in any plane (due to the rotational symmetry of the problem). 

Increasing $\kappa_0 > 0,$ the initial buckling occurs at smaller values of the forcing. The whirling behaviour still exists but is no longer symmetric, as shown in Figure \ref{fig:curvature_phasediagram_3D}b and Supplemental Video 1. The frequencies of the whirling solutions are plotted in Appendix \ref{appen:frequencies}. In fact, the asymmetry is strongest as we increase the intrinsic curvature, and choose forcing values closer to the initial bifurcation. Increasing the forcing has the effect of smoothing out the asymmetry, as the follower force dominates over the torques maintaining the intrinsic curvature.

Increasing $f$ further, we observe that whirling gives way to a new periodic solution for $\kappa_0 > 0$, which we call P1. This consists of a bent-over beating, where the filament bends in the plane it would deform in the absence of the follower force, and beats in the plane perpendicular to it, as shown in Figure \ref{fig:curvature_phasediagram_3D}a inset and Supplemental Video 2. The intrinsic curvature breaks left-right symmetry, in such a way that the filament beats in the left half of the plane. In Figure \ref{fig:curvature_phasediagram_3D}c, we fix $f=200$ and increase $\kappa_0$ from $0$ to $3\pi/4$ to show how P1 continues from the planar beating solution.  We see that increasing $\kappa_0$ increases the maximum perpendicular distance of the tip from the beat plane. We plot the frequencies of P1 for various $f$ and $\kappa_0$ in Appendix \ref{appen:frequencies}. Varying $\kappa_0>0$, planar beating is no longer visible with IVPs alone - it appears that the 2D asymmetric beating patterns observed in the planar case are unstable in the 3D model.

For larger values of $\kappa_0$, we note that the whirling state is also no longer visible using IVPs, regardless of the initial filament configuration. Furthermore, for large $\kappa_0$ and $f,$ we see a new periodic mode appearing, which we call P2. P2 beats similarly to P1, but traces a distinct shape of 8 with its tip over one period (see Figure \ref{fig:curvature_phasediagram_3D}a, inset).

\subsection{Bifurcation and Stability Analyses}

\begin{figure}
    \centering
    \includegraphics[width=0.6\linewidth]{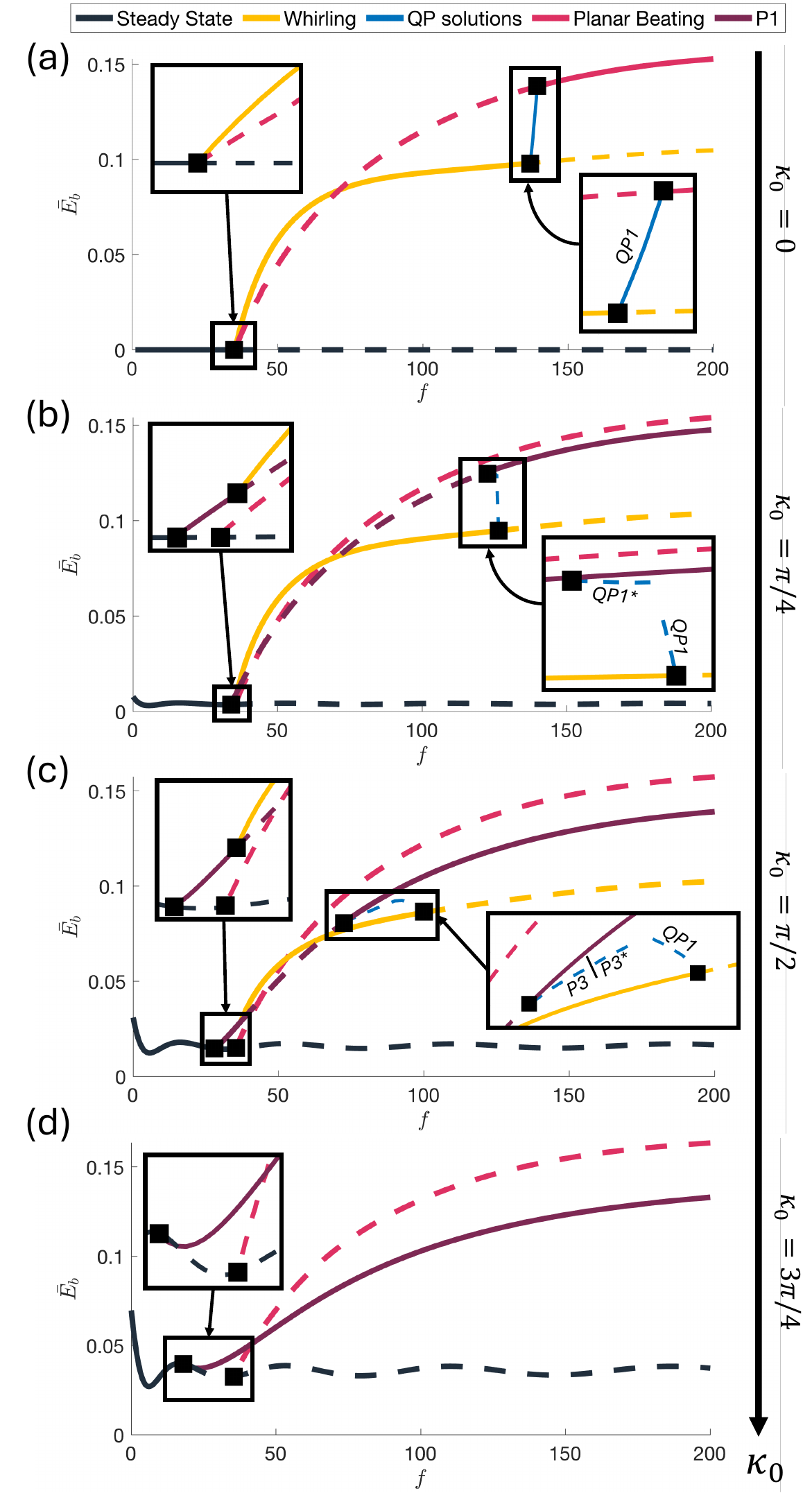}
    \caption{Bifurcation diagrams showing the different solutions as we vary $f$, and the corresponding mean bending energy of these solutions, $\bar{E}_b$, for an isotropic filament ($\beta = 1$) with intrinsic curvature for (a) $\kappa_0 = 0$, (b) $\kappa_0 = \pi/4$, (c) $\kappa_0 = \pi/2$ and (d) $\kappa_0 = 3\pi/4$. Dashed/full lines refer to unstable/stable solutions. Square markers are used to indicate the location of bifurcations as we vary $f$. Inset, left, shows the bifurcation diagram near the initial buckling event and inset, right, shows the bifurcation diagram around the bistable regions, if these are present. As $\kappa_0$ increases we see the emergence of the P1 solution branch (b-d), from which whirling bifurcates (b and c, left inset). We observe bistability between P1 and whirling (b and c, right inset) and, for the largest values of $\kappa_0$, whirling vanishes completely (d).}
    \label{fig:curvature_bifurcation_diagrams}
\end{figure}

\begin{figure}
    \centering
    \includegraphics[width=0.8\linewidth]{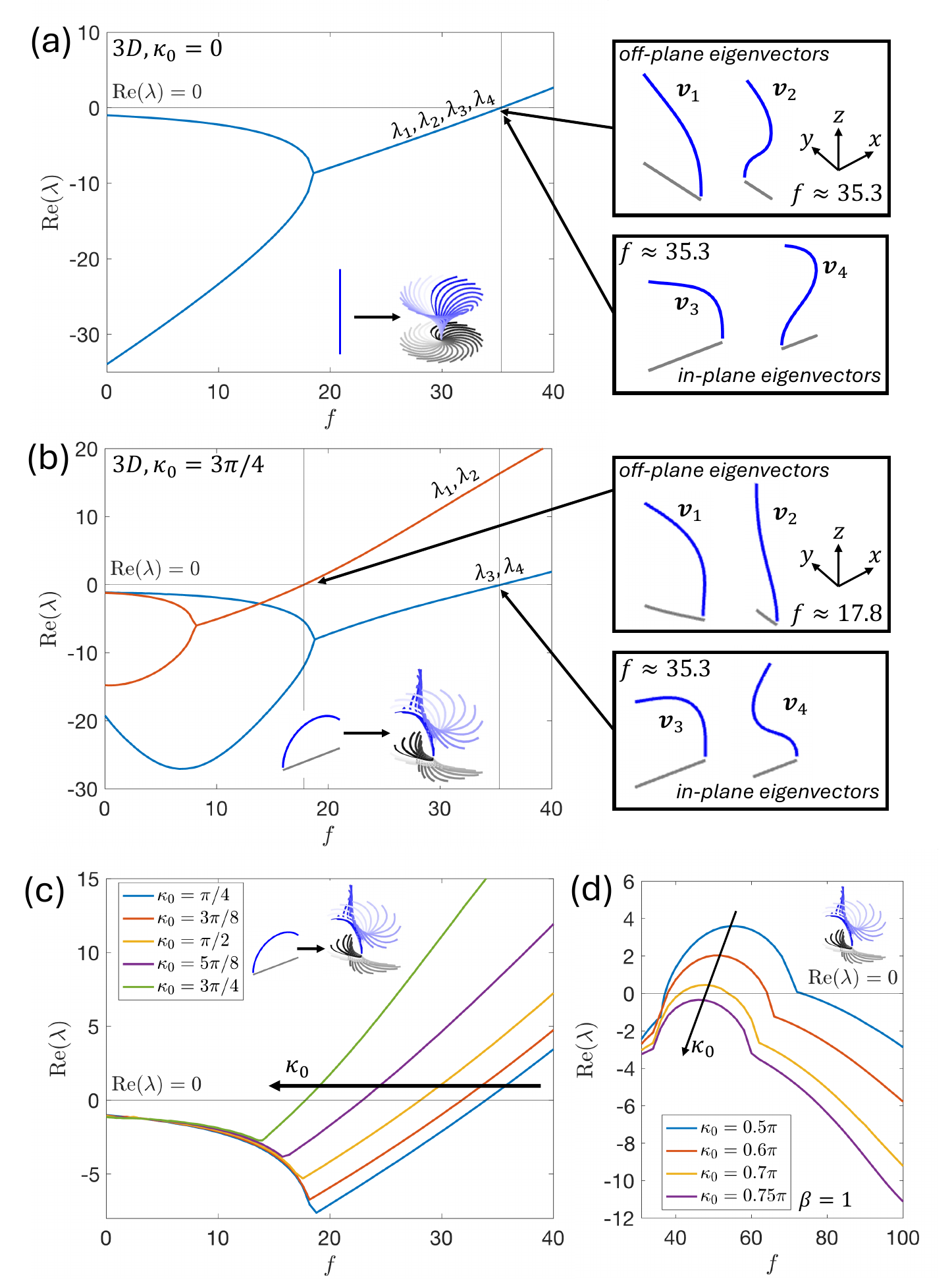}
    \caption{The real part of the four largest eigenvalues, $\lambda_i$ for $i=1,...,4$, associated to the steady state for $\beta=1$, (a) $\kappa_0 = 0$, and (b) $\kappa_0 = 3\pi/4$. Analysing the corresponding eigenvectors at each bifurcation, obtained as described in the main text and displayed on the right, allows us to identify the branches corresponding to off-plane and in-plane beating. (c) The real part of the largest eigenvalue associated to the steady state for various values of $\kappa_0.$ We see that the eigenvalue becomes unstable for smaller $f$ as we increase $\kappa_0.$ (d) The real part of the largest eigenvalue associated to the P1 state for various values of $\kappa_0$. For $\kappa_0 > \kappa_0^*$, we see the eigenvalue remains stable. }
    \label{fig:curvature_stability}
\end{figure}

To explore in more detail how filament dynamics change with $\kappa_0,$ we track the steady and time-periodic states with JFNK and perform a stability analysis to establish bifurcations within the system. An overview of the resulting bifurcation diagrams for $f\in[0,200]$ is shown in Figure \ref{fig:curvature_bifurcation_diagrams}, where we plot the non-dimensionalised mean bending energy, given by
\begin{equation}
    \bar{E}_b = \frac{1}{T} \int_0^T  \int_0^L \frac{1}{2} \kappa(s)^2 \;ds  \;dt,
\end{equation}
for each solution branch for various values of $\kappa_0$. Here $\kappa(s)$ is the local curvature of the filament. While the bending energy is usually defined with respect to the intrinsic curvature, we instead use the standard bending energy described above for ease of comparison as we vary $\kappa_0$. We now discuss the key changes in the bifurcation diagrams as we increase $\kappa_0$.  We provide a more detailed version of this discussion in Appendix \ref{appen:kappa0_bifurcations}.

\subsubsection{Isotropic Case}
We begin by recapping the isotropic case, where $\kappa_0 = 0$, shown in Figure \ref{fig:curvature_bifurcation_diagrams}a. Focusing first on the initial buckling event, in 
\citep{Clarke2024BifurcationsFilaments,Schnitzer2025OnsetFilament} it was shown that two complex conjugate pairs of eigenmodes become unstable at $f=f^*$ (see Figure \ref{fig:curvature_stability}a). We hence identify this bifurcation as a Double Hopf bifurcation. 

To understand the buckling bifurcation further, we visualise the eigenvectors. Defining each eigenvalue-eigenvector pair as $(\lambda_i, \bm{x}_i)$, we note that the $\bm{x}_i$ are complex for $i=1,...,4$. Therefore, to interpret their physical meaning, we take linear combinations of the eigenvectors, $\bm{v}_1 = a(\bm{x}_1 + \bm{x}_2), \bm{v}_2 = a(\bm{x_1} - \bm{x}_2)$ and similar for $\bm{v}_3$ and $\bm{v}_4$ using $\bm{x}_3$ and $\bm{x}_4$, where $a$ is an arbitrary constant chosen to emphasise the structure of the eigenmode. As the magnitude of the state variable relates to angles of rotation of the local basis, taking larger values of $a$ results in more exaggerated filament deformation. We show these eigenvectors in Figure \ref{fig:curvature_stability} for $a=5.$ We observe that the first complex conjugate pair is associated to beating in the $(x,z)-$plane, and the second is associated to beating in the perpendicular plane, as shown by the eigenvectors in Figure \ref{fig:curvature_stability}a (right). It is these modes that combine to give rise to the stable whirling solution observed after the bifurcation, as well as the unstable planar beating branch. For larger $f$, whirling becomes unstable and gives rise to QP1. The QP1 branch exists for a small region of $f$, and vanishes when planar beating stabilises for larger $f$ (see Figure \ref{fig:curvature_bifurcation_diagrams}a, inset). Further discussion on the bifurcation diagram in the isotropic case, including exploration of higher forcing values, can be found in \citep{Clarke2024BifurcationsFilaments}.

\subsubsection{Initial Buckling Event}
As we increase $\kappa_0$, we see a change in the initial buckling event; as shown in Figure \ref{fig:curvature_bifurcation_diagrams}a-d left inset, the initial buckling happens at smaller values of $f$ as we increase $\kappa_0.$ We also now observe two solution branches bifurcating from the steady state at different values of the forcing, and note that whirling no longer bifurcates from the steady state for $\kappa_0>0$. Instead, we find the P1 branch bifurcates from the steady state initially at a critical forcing $f<f^*$, followed by the unstable planar beating branch at $f\approx f^*$. To explore these changes in the bifurcation diagrams, we analyse the eigenmodes associated to the steady state. 

We plot the real parts of the four dominant eigenmodes of the steady state for $\kappa_0 = 3\pi/4$ in Figure \ref{fig:curvature_stability}b, including visualisations of the eigenvectors obtained as discussed previously. We see that the two modes corresponding to buckling in either plane separate; the mode corresponding to in-plane, asymmetric beating still becomes unstable at $f\approx f^*,$ whereas the second mode, corresponding to off-plane beating, becomes unstable earlier. We note that the off-plane eigenmodes do not both lie in the $(y,z)-$plane. This is because the intrinsic curvature breaks the symmetry about the $(y,z)$-plane, and so we expect one of these eigenmodes to reflect this broken symmetry. This can be seen in the off-plane eigenvectors (see Figure \ref{fig:curvature_stability}b, top right). Hence for $\kappa_0>0$, at the initial bifurcation, it is only the off-plane modes that are becoming unstable which lead to P1 through a Hopf bifurcation. In Figure \ref{fig:curvature_stability}c, we plot the real part of the dominant eigenmode for various values of $\kappa_0$. We see that the off-plane modes become unstable earlier as we increase the intrinsic curvature, meaning the steady state buckles at smaller forcing values, as we observed earlier by solving IVPs. 

The off-plane eigenmodes destabilising for lower $f$ as we increase $\kappa_0$ is a result of the intrinsic curvature being defined in the filament's local frame. Consequently, 3D perturbations to the filament, which lead to perturbations in the filament frame, result in 3D perturbations to the intrinsic curvature. The off-plane components of the intrinsic curvature perturbations result in off-plane internal moments, leading to the initial buckling occurring at smaller $f$. Further investigation and discussion of the dependence of initial buckling on the intrinsic curvature can be found in Appendix \ref{appen:initialbuckling_alpha}.

Immediately following the initial buckling bifurcation for $\kappa_0>0$, we see from Figures \ref{fig:curvature_bifurcation_diagrams}b-d that the P1 solution is stable. For moderate $\kappa_0$, the P1 solution becomes unstable for larger $f$, giving way to the stable whirling solution (see Figures \ref{fig:curvature_bifurcation_diagrams}(b-c), left inset). This can again be linked to the initial buckling event. As the whirling solution is known to arise from a combination of both the in-plane and off-plane buckling eigenmodes \citep{Clarke2024BifurcationsFilaments}, it follows that the whirling solution is always born after the second buckling mode becomes unstable, branching from the P1 solution branch.

\subsubsection{Bistability}
Recall that for $\kappa_0=0$, we observe a stable branch (QP1) connecting the whirling and beating solutions. As we increase $\kappa_0$, we see that the stable branch connecting P1 and whirling exists for smaller ranges of $f$, until it is no longer stable. In this case, i.e. for moderate $\kappa_0,$ we observe bistability between the P1 and whirling solutions (see Figures \ref{fig:curvature_bifurcation_diagrams}b,c). As such, we use bisection to unearth the unstable solution branches which exist in the region. We find multiple distinct solutions, including the same QP1 solution from the isotropic case, shown in Figure \ref{fig:curvature_bifurcation_diagrams}b,c right inset. The sharp turns in the bifurcation diagrams indicate there could be further bifurcations occuring in this region, however we are unable to identify these using JFNK or bisection alone. We discuss these unstable solution branches further, including plots of their dynamics, in Appendix \ref{appen:kappa0_bifurcations}.

\subsubsection{Loss of Whirling Solution}
In the previous sections we identify three key bifurcations; P1 becoming unstable, P1 subsequently restabilising, and whirling becoming unstable. As we increase $\kappa_0$, these three bifurcations occur within a smaller range of $f$, until the intrinsic curvature reaches a critical value, and the bifurcations coaelse. For larger $\kappa_0,$ for instance $\kappa_0=3\pi/4$ as shown in Figure \ref{fig:curvature_bifurcation_diagrams}d, the whirling solution vanishes completely, and we only observe the stable P1 solution branch.

We can explore the disappearance of the whirling solution by performing Floquet analysis on the P1 solution for various values of $\kappa_0$. The real part of the dominant eigenvalue for each $\kappa_0$ is shown in Figure \ref{fig:curvature_stability}d. We see that the two bifurcations, corresponding to P1 becoming unstable and then restabilising, get closer together as we increase $\kappa_0$, until they merge at a critical value, $\kappa_0=\kappa_0^*$. At this value, the whirling solution collapses onto the P1 branch, also terminating QP1 and QP1* in the process. Beyond this $\kappa_0,$ only P1 is stable in this region of $f.$ We can identify the loss of the whirling solution by again revisiting the initial buckling event. For $\kappa_0 > \kappa_0^*$, it appears that the growth rate associated to the off-plane modes is so large compared to the growth rate of the in-plane modes that the whirling solution never exists. Instead, we only observe P1, matching observations from IVPs.

%\begin{figure}
%    \centering
%    \includegraphics[width=0.5\linewidth]{curvature_bifurcation_diagrams.pdf}
%    \caption{Bifurcation diagrams for a filament with intrinsic curvature given by (a) $\kappa_0 = 0$, i.e. no curvature, (b) $\kappa_0 = \pi/4$, (c) $\kappa_0 = \pi/2$, and (d) $\kappa_0 = 3\pi/4$. Dotted (/full) lines refer to unstable (/stable) solutions.}
%    \label{fig:curvature_bifurcation_diagrams}
%\end{figure}
%\begin{figure}
%    \centering
%    \includegraphics[width=0.8\linewidth]{curvature_stability.pdf}
%    \caption{Stability analyses for a filament with intrinsic curvature $(\beta=1)$. (a) The real part of the dominant four eigenvalues for $\kappa_0=3\pi/4$, corresponding to in-plane and off-plane buckling. Right: eigenvectors corresponding to off-plane (top) and in-plane (bottom) branches at $f=20$. (b) The real part of the dominant eigenvalue of the steady state for various values of $\kappa_0$. Inset: snapshots of the corresponding steady state at the critical forcing value for various $\kappa_0$, found via JFNK. (c) Real part of the dominant Floquet exponent associated to P1 for various values of $\kappa_0$.}
%    \label{fig:curvature_stability}
%\end{figure}
\section{Aniostropic filaments with no intrinsic curvature}

\subsection{Initial Value Problems}
\begin{figure}
    \centering
    \includegraphics[width=0.75\linewidth]{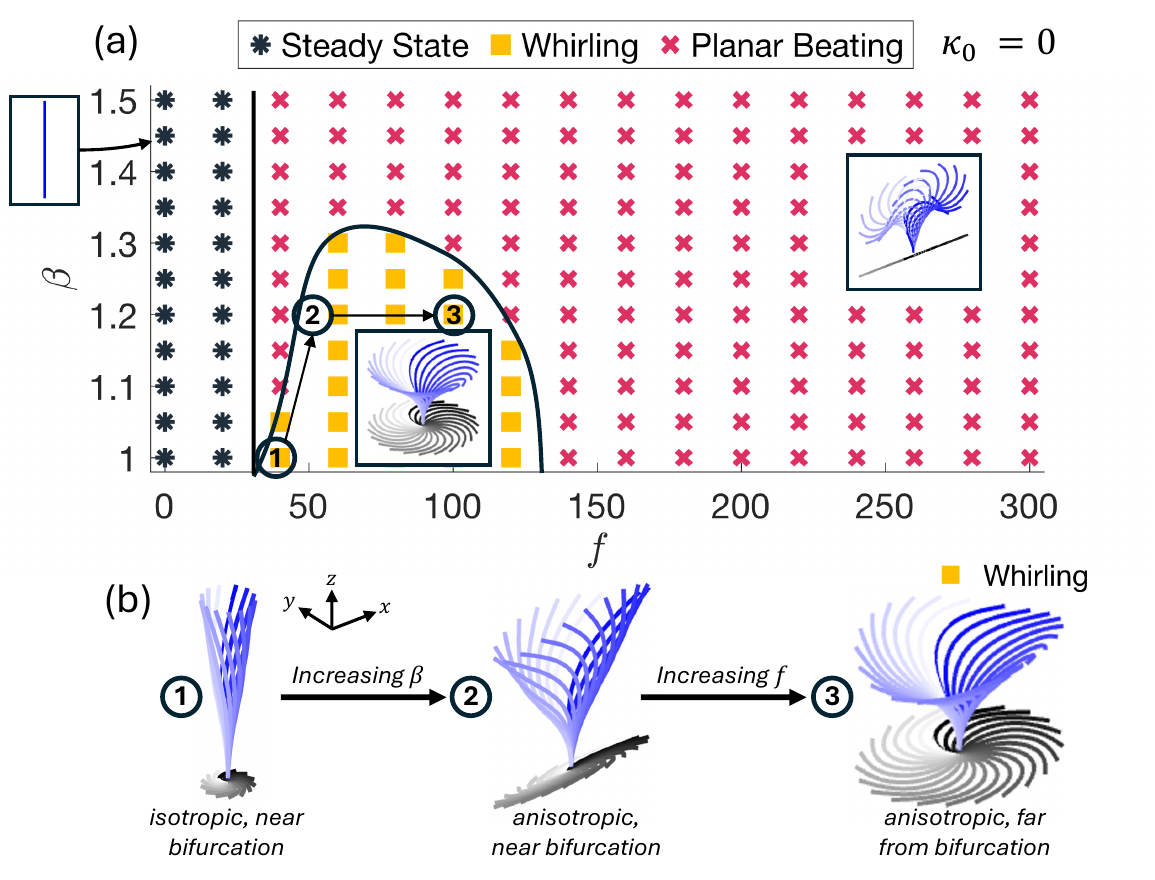}
    \caption{(a)  Results of numerical simulations of an anisotropic filament with no intrinsic curvature $(\kappa_0 = 0, \beta \geq 1)$. The black lines are drawn to indicate the approximate boundaries of the solution regions. (b) Snapshots of filaments over one period undergoing whirling dynamics in the isotropic case, close to the bifurcation $(\beta = 1, f = 36)$, an anisotropic case near the bifurcation $ (\beta=1.2, f=46)$ and an anisotropic case further from the bifurcation $(\beta = 1.2, f=100)$. The three dynamics are indicated on the phase diagram in (a) using numbered circles.}
    \label{fig:bending_phasediagram}
\end{figure}

In the previous section we investigated the effect of intrinsic curvature defined in the plane defined by $\hat{\bm{\nu}}$ and $\hat{\bm{t}}$.  Now, we set $\kappa_0 = 0$ and focus on the effect of introducing anisotropy through the bending stiffness by introducing the parameter $\beta$, which describes the ratio of the bending stiffnesses about the $\hat{\bm{\nu}}-$ and $\hat{\bm{\mu}}-$axes. Throughout this paper we focus on $\beta>1,$ but the $\beta<1$ case follows immediately by considering a rotation of the axes, and redefining the non-dimensional follower force as $\tilde{f} = \beta f$ (c.f. Eq \eqref{eq:followerforce_BC}). In Figure \ref{fig:bending_phasediagram}a, we show the results of IVPs for $\beta \in [1,1.5]$ and $f\in [0,300].$ Regardless of $\beta$ and $f$, the steady state is the vertically upright filament. For $\beta=1$, this buckles into whirling, while for $\beta > 1$ the steady state buckles into planar beating in the $(x,z)-$plane. For moderate values of $\beta$, beating becomes unstable as we increase the forcing, with whirling appearing in its place. The frequency of these whirling solutions for different $\beta$ and $f$ values can be found in Appendix \ref{appen:frequencies}. Close to the bifurcation in which planar beating appears to become unstable, whirling is very similar to in-plane beating, with a small off-plane component. This causes the tip to trace an ellipse with a major axis that aligns with the $\hat{\bm{x}}-$axis, as shown in Figure \ref{fig:bending_phasediagram}b, middle and Supplemental Video 3. The large amplitude in the $\hat{\bm{x}}-$direction close to the bifurcation suggests that we can expect the whirling solution to branch from the planar beating solution branch, instead of the steady state. Further from the bifurcation, however, the off-plane component grows and whirling resembles the isotropic case more closely (see Figure \ref{fig:bending_phasediagram}b, right). For larger forcing values, whirling becomes unstable and planar beating in the $(x,z)-$plane reappears. As we increase $\beta$, the region in which whirling is stable appears to decrease, until $\beta = 1.35$, where we no longer observe whirling through IVPs. For these $\beta$ values, we only observe planar beating in the $(x,z)-$plane after buckling.

\begin{figure}
    \centering
    \includegraphics[width=\linewidth]{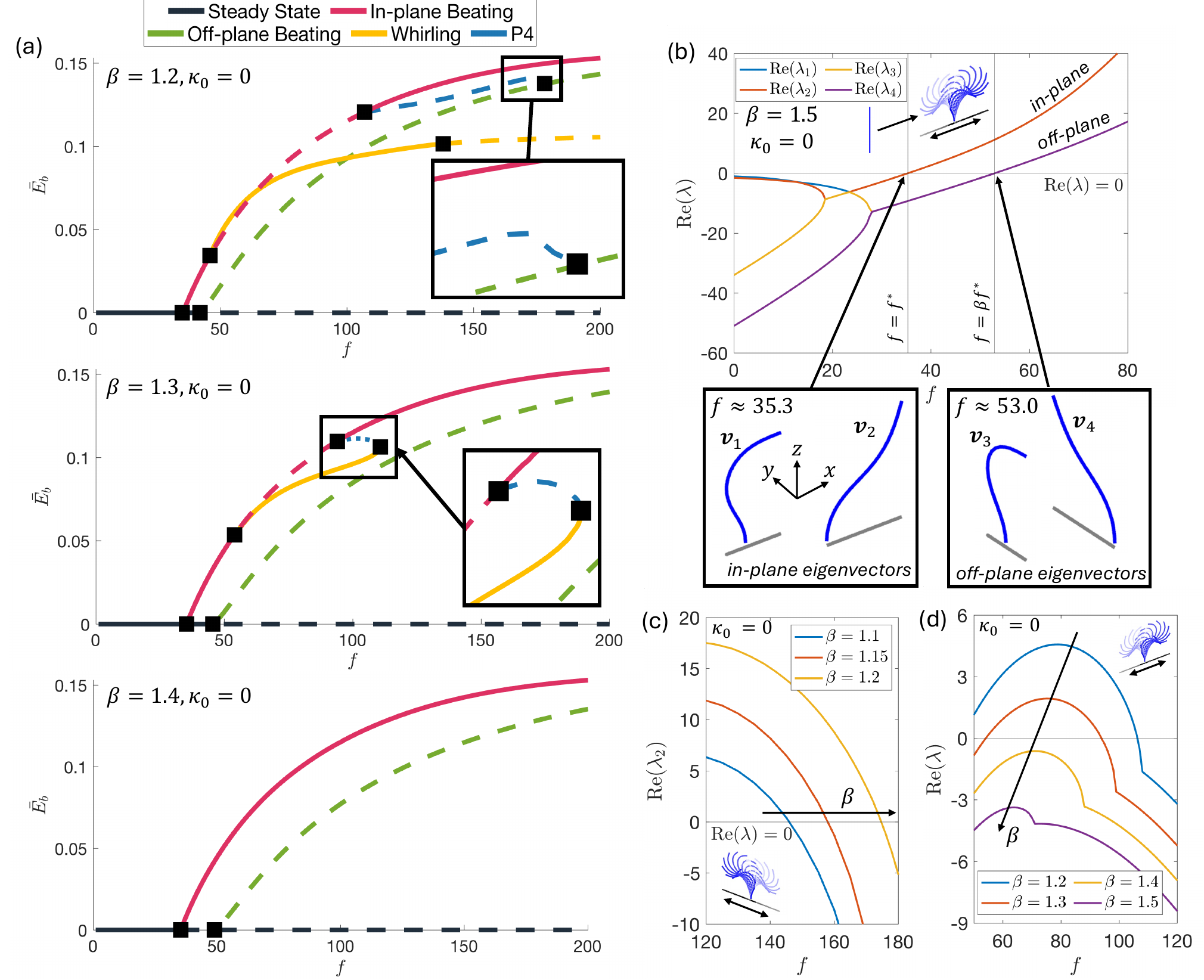}    
    \caption{(a) Bifurcation diagrams showing the different solutions as we vary $f$, and the corresponding mean bending energy of these solutions, $\bar{E}_b$, for an anisotropic filament with no intrinsic curvature ($\kappa_0 = 0$) for $\beta = 1.2, 1.3$ and $1.4$ (top to bottom). Dashed/full lines refer to unstable/stable solutions. Square markers are used to indicate the location of bifurcations as we vary $f$.
    We see the separation of the in-plane and off-plane solution branches as $\beta$ increases, and observe bistability between in-plane beating and whirling (inset). For larger values of $\beta$, the whirling solution branch vanishes completely (bottom).
    (b) The real part of the four largest eigenvalues associated to the steady state for $\beta = 1.5$. As described in the main text, we visualise the eigenvectors, $\bm{x}_i$ for $i=1,...,4$, by taking linear combinations of the complex conjugate pairs, for example $\bm{v}_1 = a(\bm{x}_1 + \bm{x}_2)$ where $a$ is a constant. We plot the eigenvectors for $a=5$ to identify the branches corresponding to in-plane beating (bottom left), and off-plane beating (bottom right). (c) The real part of the second largest eigenvalue associated to off-plane beating for various values of $\beta.$ We associate this eigenvalue becoming stable with P4 collapsing onto the off-plane beating branch. (d) The real part of the largest eigenvalue associated to the in-plane beating state for various values of $\beta$. For $\beta > \beta^*$, we see the eigenvalue remains stable. }
    \label{fig:bending_bifurcation_diagrams_and_stability}
\end{figure}

\subsection{Bifurcation and Stability Analyses}
Using JFNK to track the steady and periodic solutions and then performing a stability analysis on the subsequent branches allows us to generate the bifurcation diagrams shown in Figure \ref{fig:bending_bifurcation_diagrams_and_stability}a. We outline the key features of the bifurcation diagrams below, and provide further details in Appendix \ref{appen:beta_bifurcations}.

Focusing first on the initial bifurcation, we see that the critical forcing value at which the steady state buckles remains at $f=f^*$ regardless of $\beta$. At this point, the planar beating branch in the $(x,z)-$plane, which we will henceforth call `in-plane' beating (opposed to `off-plane' beating, in the $(y,z)-$plane), is born through a Hopf bifurcation. We can also identify the unstable off-plane beating branch for each $\beta$, which also appears to connect to the trivial steady state, bifurcating at $f= \beta f^*$. These solutions are present regardless of $\beta$, as shown in Figure \ref{fig:bending_bifurcation_diagrams_and_stability}a. These branches are linked to the initial buckling event; introducing $\beta \neq 1$ has the effect of separating the critical values at which the beating in either plane becomes unstable, similar to the effect of having $\kappa_0 \neq 0$. This is a consequence of the non-dimensionalisation of the follower force, as given in Eq \eqref{eq:followerforce_BC}. Accordingly, we expect eigenmodes to become unstable at $f=f^*$ (corresponding to buckling in the $(x,z)-$plane) and $f = \beta f^*$ (corresponding to buckling in the $(y,z)-$plane). Thus, the magnitude of $\beta$ dictates which plane we initially buckle into, and furthermore when successive buckling modes become activated. To verify the effect of $\beta$ on the initial buckling bifurcation, we plot the dominant eigenvalues for $\beta=1.5$ in Figure \ref{fig:bending_bifurcation_diagrams_and_stability}b. As expected, we find that the first bifurcation occurs at $f= f^*,$ with corresponding planar eigenmodes lying in the $(x,z)-$plane. Therefore, the filament is expected to undergo planar deformations in this plane, as is confirmed by the numerical simulations. The second mode, corresponding to buckling in the perpendicular plane, is found to occur at $f =  \beta f^* \approx 53.0$. 

For moderate $\beta$ (see Figures \ref{fig:bending_bifurcation_diagrams_and_stability}a top, middle) we find that, for a larger value of the forcing, in-plane beating becomes unstable and the whirling solution is born. Between the bifurcations in which in-plane beating becomes stable again and whirling becomes unstable, there is a region of bistability. Bisection in this region yields a single solution branch, which we call P4. As P4 is periodic, we can track the behaviour using JFNK. In doing so, we find that P4 connects to the off-plane beating branch, rather than the whirling branch, as shown in Figure \ref{fig:bending_bifurcation_diagrams_and_stability}a (top, inset). Floquet analysis of the off-plane beating branch confirms this, showing that the second largest eigenvalue, $\lambda_2$, becomes stable when P4 collapses onto the branch. By performing a Floquet analysis on the off-plane beating branch for various values of $\beta,$ as shown in Figure \ref{fig:bending_bifurcation_diagrams_and_stability}c, we see that $\lambda_2$, becomes stable at higher $f$ as we increase $\beta$, indicating that P4 joins the off-plane beating branch at higher values of $f$ as we increase $\beta$. This trend continues until a critical value of $\beta,$ beyond which P4 now connects to the whirling branch, after which both branches terminate. Consequently, for these values of $\beta,$ the whirling branch only exists when it is stable (see Figure \ref{fig:bending_bifurcation_diagrams_and_stability}a middle).

The three bifurcation points corresponding to in-plane beating becoming unstable, becoming stable again, and the whirling branch disappearing all occur within smaller ranges of $f$ as we increase $\beta$, until the three bifurcations collide and we see only one stable branch - in-plane beating, as shown in Figure \ref{fig:bending_bifurcation_diagrams_and_stability}a (bottom) for $\beta = 1.4.$ Indeed, a stability analysis of the in-plane beating solution, shown in Figure \ref{fig:bending_bifurcation_diagrams_and_stability}d, confirms that the range of $f$ for which beating is unstable decreases with increasing $\beta,$ up to a critical value of $\beta = \beta^*$ where the two pitchfork bifurcations collide, which likely also coalesces with the P4-whirling bifurcation. Beyond this, beating is stable for all values of $f$ in our tested range and, as in the intrinsic curvature case, whirling ceases to exist entirely. Both the Floquet analysis and IVPs suggest that the whirling branch collapses onto the beating branch at $\beta=\beta^*,$ resulting in the loss of P4 and whirling as distinct solutions. The disappearance of whirling can again be explained through the large separation in critical force values between in-plane and off-plane beating solutions bifurcating from the trivial state.

\section{Using bending stiffness anisotropy to stabilise asymmetric beating}
%\begin{itemize}
    %\item Focus on $\kappa_0=3\pi/4$
    %\item Initial buckling has competing effects from $\kappa_0$ and $\beta$
    %\item Increasing $\beta$ births whirling again
    %\item Here, observe a bunch of bistability as planar beating and P1 compete
    %\item Whirling dies again above a critical value of $\beta$ (though P1 never dies, comes from inital buckling)
    %\item Complete collapse onto 2D happens for larger $\beta$. For $\beta=1.5,$ IVPs indicate collapse has occured, but floquet shows P1 and PB are bistable for a region. By $\beta=2$, P1 is always unstable. (Discuss here: beating asymmetry more pronounced close to bifurcation)
%\end{itemize}

\begin{figure}
    \centering
    \includegraphics[width=0.8\linewidth]{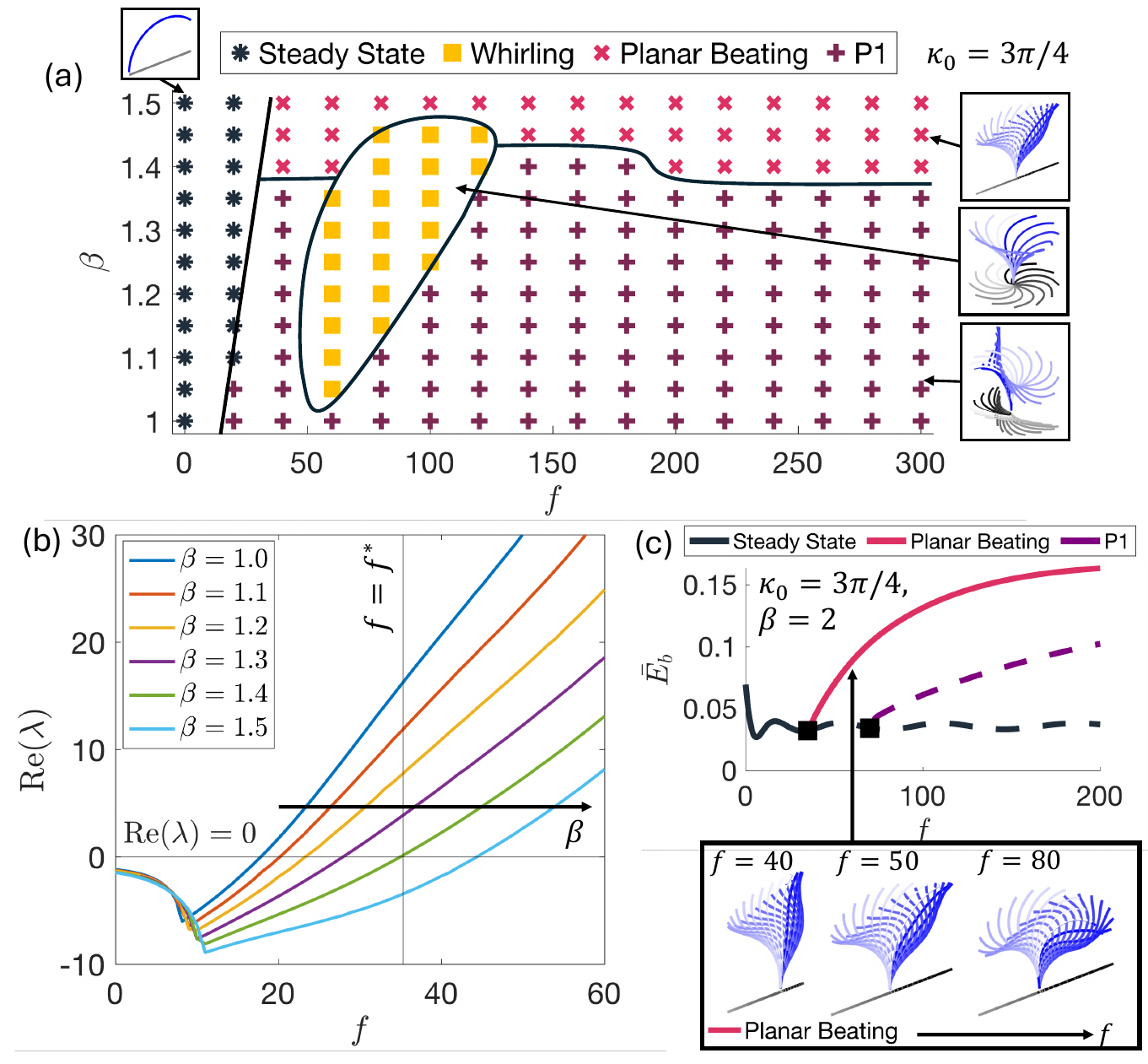}
    \caption{(a) Results of numerical simulations of an anisotropic filament with intrinsic curvature $(\kappa_0 = 3\pi/4, \beta \geq 1)$. The black lines are drawn to indicate the approximate boundaries of the solution regions. (b) The eigenvalue associated to off-plane beating for several values of $\beta$. For $\beta \lesssim 1.4$, this eigenmode becomes unstable before the in-plane eigenmode (i.e. before $f=f^*$). For $\beta \gtrsim 1.4$, the in-plane eigenmode becomes unstable first. (c) The bifurcation diagram showing the different solutions as we vary $f$, and the corresponding mean bending energy of these solutions, $\bar{E}_b$, for $\kappa_0=3\pi/4$, $\beta=2.$ We see that only asymmetric planar beating (bottom) is stable. }
    \label{fig:alpha_and_beta}
\end{figure}

Following the previous section, we see that increasing $\beta>1$ has the effect of stabilising planar solutions. Thus we now combine the effects of $\kappa_0 \neq 0$ and $\beta \neq 1$, with the goal of recovering the planar asymmetric beating observed in 2D simulations of filaments with intrinsic curvature, which were found to be unstable to 3D perturbations in Section \ref{sec:curvature} for $\beta = 1$. Both the P1 and whirling states must be eradicated for a collapse onto planar dynamics. Following the results of the previous sections, we anticipate that the stability of the P1 branch and existence of whirling solution are related to the eigenmodes of the initial buckling event. As increasing $\kappa_0$ appears to cause an earlier buckling of the off-plane eigenmodes, while increasing $\beta$ delays the onset of this instability, we expect that increasing the intrinsic curvature will require a larger amount of anisotropy to stabilise the planar modes. As such, we focus on $\kappa_0=3\pi/4$, anticipating this to provide an upper bound for the value of $\beta$ which causes a collapse to 2D dynamics, and increase $\beta$ from $\beta = 1$ to $\beta = 1.5.$ The results from IVPs over this parameter range are shown in Figure \ref{fig:alpha_and_beta}a.

As discussed in previous sections, $\kappa_0$ and $\beta$ have opposite effects on the initial buckling modes; increasing the intrinsic curvature (i.e. increasing $\kappa_0$) causes the off-plane modes to buckle earlier, whereas increasing the anisotropy through the bending stiffness ratio (i.e. increasing $\beta$) causes later buckling of these modes. In both cases, in-plane buckling happens at the same critical forcing, regardless of $\kappa_0$ or $\beta$. Therefore, fixing $\kappa_0$ and increasing $\beta$ shifts the off-plane beating modes to buckle at higher values of the forcing. This is shown in Figure \ref{fig:alpha_and_beta}b, which displays the eigenvalues corresponding to off-plane buckling for various values of $\beta$. As we can see, at $\beta \approx 1.4$ the off-plane buckling occurs at $f\approx f^*$, the critical forcing at which in-plane buckling occurs. Beyond this $\beta,$ the off-plane buckling occurs for $f>f^*$, indicating that the initial buckling of the filament will be to in-plane beating, and will occur at $f\approx f^*.$ Accordingly, as we increase $\beta$ from $1$, the IVPs in Figure \ref{fig:alpha_and_beta}a show that the initial buckling into P1 happens at a critical value of the forcing which increases with $\beta$ until it reaches $f\approx f^*$ at $\beta \approx 1.4,$ beyond which the filament buckles into in-plane asymmetric beating at $f\approx f^*$.

Furthermore, as increasing $\beta$ increases the critical forcing value at which off-plane beating bifurcates from the steady state, increasing the bending stiffness anisotropy results in the birth of whirling again (as illustrated by the IVPs in Figure \ref{fig:alpha_and_beta}a). For these intermediate values of $\beta,$ Floquet analysis indicates bistability between P1, whirling and in-plane beating at several points, likely linked to the existence of distinct unstable quasiperiodic and periodic solutions joining the three branches. However as $\beta$ is increased further, and the off-plane buckling modes become unstable for larger $f,$ we reach a scenario where the whirling branch dies again above a critical value of $\beta$. This occurs through the same mechanism as the $\kappa_0=0$ and $\beta=1$ cases.

As expected, increasing the bending stiffness anisotropy further leads to a complete collapse onto planar dynamics. While initial value problems from the upright-state indicate that this collapse has occurred for $\beta=1.5$ (see Figure \ref{fig:alpha_and_beta}a), Floquet analysis shows that there are still regions where P1 is stable for this choice of parameters, alongside in-plane beating. By $\beta=2$, however only in-plane beating is stable, showcasing the same asymmetric beating dynamics as the planar case for $\kappa_0=3\pi/4$, with particular asymmetry close to the bifurcation. We show the bifurcation diagram for $\beta = 2,$ and the planar beating dynamics, in Figure \ref{fig:alpha_and_beta}c  and in Supplemental Video 4.

\section{Summary and Discussion}

%\begin{itemize}
    %\item Key results: loss of double Hopf, whirling vanishes
    %\item Preferred curvature used in the literature to get cilia-like beats, but this is unstable in 3D
    %\item Values of $\kappa_0$ and $\beta$ used in other work/in the literature
    %\item Respiratory cilia have full axoneme, so preferential bending due to 5-6 bridge and due to central pair, and undergo almost planar beating (asymmetric). Nodal cilia don't have central pair, and undergo a bent-over whirling
    %\item Future work - experimentally, predicting the values of $\beta$. Computationally, incorporating these with more accurate forcing mechanisms, but also looking at collective effects of cilia
    %\item Future work (more specific)/discussion: QP solutions/unstable solutions between whirling and P1/PB aren't fully explored, as we can't track with JFNK. Bisection implies an introduction of new QP solutions when asymmetry/anisotropy is introduced, as a pathway to the new, asymmetric/anisotropic beating. Potential of extra bifurcations occuring between these behaviours - need to explore further. for anisotropy, don't know if P3 is QP1, or if there is another behaviour - evals of perp PB don't imply so, but we can't use bisection to find it
%\end{itemize}

In this paper, we investigated the effects of asymmetry and structural anisotropy in a follower-force driven filament, through the introduction of both intrinsic curvature and bending stiffness anisotropy.  In both cases, we observed that the initial buckling changed from a double Hopf bifurcation, which is observed in the symmetric case 
\citep{Clarke2024BifurcationsFilaments,Schnitzer2025OnsetFilament}, to a single Hopf, suggesting that the double Hopf is a special case of these systems and is related to symmetry. When varying only the intrinsic curvature, in 2D simulations we obtain the same asymmmetric beating reported in the literature \citep{Wang2023Generalized-NewtonianFilament}. However in 3D, we see that this beating is unstable, with the filament instead buckling into a fully 3D beating occurring primarily perpendicular to the plane the filament would deform in the absence of the follower force (P1). Increasing the forcing, we observe a tilted whirling and discuss how this changes with both $f$ and $\kappa_0$. For moderate curvatures, we observe that P1 and whirling become bistable, making the stable QP1 branch connecting whirling to beating in the zero-curvature case now unstable, and introducing further unstable solutions which we identify through bisection. Similar results hold when varying the bending stiffness anisotropy, replacing P1 with planar beating in the plane of preferential bending. In both cases, we observe that as we increase the amount of asymmetry, whirling does not only become unstable, but rather vanishes completely as a solution. Instead, we observe only P1 for the intrinsic curvature case, or in-plane beating for the anisotropic bending stiffness case. Finally, we show how introducing anisotropy through the bending stiffness on a filament with intrinsic curvature can re-introduce whirling as a solution for moderate $\beta$ values, and ultimately stabilise the planar asymmetric beating observed in 2D for isotropic filaments with intrinsic curvature for larger $\beta$ values. We find that a difference by a factor of 2 in the bending stiffnesses is sufficient for this collapse into planar beating for the highest value of $\kappa_0$ tested for the follower force model; a remarkably reasonable ratio.

Changing the intrinsic curvature has the effect of adding asymmetry into the beating pattern, motivating its inclusion in recent numerical studies \citep{Chakrabarti2019SpontaneousMicrofilaments,Chakrabarti2019HydrodynamicFilaments,Chakrabarti2022ACilia,Sartori2016DynamicFlagella,Cass2023TheFlagella,Wang2023Generalized-NewtonianFilament}. These 2D works use values $\kappa_0 \in [2,2.71]$ to match the beating pattern to experiments, in particular to wild-\textit{Chlamydomonas} flagella \citep{Chakrabarti2019SpontaneousMicrofilaments,Sartori2016DynamicFlagella}. However this is not the only potential cause of asymmetry in ciliary beating; some works suggest that the bending stiffness in each plane may not be constant, instead utilising a curvature dependent bending stiffness \citep{Moreau2024MinimalCilium,Han2018SpontaneousCilia}. In such a model, the in-plane bending stiffness can take one of two values depending on the sign of the local curvature. Other works suggest that the asymmetry is a result of bias in the rates at which the dynein attach and detach from the microtubules, depending on their placement in the axoneme \citep{Chakrabarti2019SpontaneousMicrofilaments,Chakrabarti2019HydrodynamicFilaments,Chakrabarti2022ACilia}. In each of these models, numerical studies result in asymmetric dynamics exhibiting distinct recovery and effective strokes. Our results indicate that these asymmetric beating solutions could be unstable in 3D, but that they could be stabilised by incorporating anisotropy to the model, for instance structural anisotropy through the bending stiffness.

Incorporating anisotropy through the bending stiffness in either direction is motivated by the internal structure of the cilium. Cilia have a preferential bending direction both due to the orientation of the central pair of microtubules \citep{Gibbons1981CiliaEukaryotes,Marshall2006Cilia:Antenna,Gilpin2020TheFlagella} and the stiffness of the 5-6 bridge \citep{Gibbons1981CiliaEukaryotes,AFZELIUS1959ElectronFixative.} which limits sliding between microtubule doublets 5 and 6 in the axoneme. The effect of the bridge is captured in our work by allowing the perpendicular stiffness to differ from the in-plane stiffness, where `in-plane' is aligned with the preferential beating plane of the cilium. Structural anisotropy was utilised in \citep{Rallabandi2022Self-sustainedFlagellum}, where it was shown that the transition from planar to whirling dynamics is delayed as we increase the anisotropy. Based on our results, we suspect that increasing the anisotropy further in this study would eradicate the 3D whirling dynamics completely, and only planar solutions would remain. It is expected that larger values of $\beta$ may be required for different forcing mechanisms.  For example, in \citep{Rallabandi2022Self-sustainedFlagellum}, values up to $\beta = 10$ were considered and 3D dynamics were still observed, while we found values as small as $\beta = 2$ were sufficient to suppress 3D dynamics entirely for the follower force model.  In \citep{Ishijima1994FlexuralFlagella}, experimental evidence on sand dollar sperm flagella reports a value of $\beta = 13.8.$ This is much higher than the threshold for entirely planar dynamics that we find, and so could explain the planar beating observed by these cells \citep{Ishijima1994FlexuralFlagella}. In \citep{Lindemann2016FunctionalFlagellum}, a wooden axonemal model supplemented with a stronger 5-6 bridge (by using a silicon adhesive between doublets 5 and 6) and a central pair was used to generate an estimate of $\beta=2.6$ for the axoneme. 

Although the follower force model differs from the internal shear-driven sliding mechanism which generates ciliary beating, the dynamics observed in this paper are reminiscent of those observed biologically. For instance nodal cilia found in the embryo are known to whirl with a tilt to the left, generating fluid flows which break the left-right symmetry in the embryo \citep{Nonaka1998RandomizationProtein,Smith2008FluidCilia,Smith2011MathematicalCilia}, similar to the asymmetric whirling observed for moderate $\kappa_0$ and $\beta$. Motile cilia on the other hand, such as respiratory cilia, undergo asymmetric near-planar beating with low-curvature power strokes, followed by distinct high-curvature recovery strokes \citep{Chilvers2000AnalysisMethods}. While we do not observe these two components of the beat with this simple model, for large enough $\beta$ and moderate $\kappa_0$, we do observe asymmetric planar beating. Structurally, these cilia are different; nodal cilia lack the central pair of MTs in the axoneme, which respiratory cilia do have. The existence of a central pair would increase the structural anisotropy, and hence increase $\beta$. This is consistent with our findings that filaments with lower $\beta$ values can access tilted whirling as a state, whereas higher $\beta$-filaments can only achieve planar beating.

The follower force model is a simple, commonly-used model that yields cilia-like dynamics through instabilities and provides an environment to explore the effect of anisotropies on these states.  Some other forcing mechanisms more accurately describe the shear-driven activation of cilia, instead considering the forcing with a multi-level approach; considering how microscale attachment and detachment of dynein motors in the axoneme can lead to large scale ciliary beating. In such models, anisotropy can be included through the bending stiffness \citep{Rallabandi2022Self-sustainedFlagellum}, or by including stiff springs to represent the 5-6 bridge in the axoneme \citep{Han2018SpontaneousCilia}. Then asymmetric beating can be achieved through the intrinsic curvature \citep{Sartori2016DynamicFlagella,Cass2023TheFlagella}, or through biased feedback mechanisms which control the forcing output of the motors \citep{Chakrabarti2019SpontaneousMicrofilaments,Chakrabarti2019HydrodynamicFilaments,Chakrabarti2022ACilia}. Exploring how anisotropies through the curvature and stiffness affect dynamics in models with more accurate forcing mechanisms would give more accurate estimates for the $\kappa_0$ and $\beta$ values necessary to recover beats closer to those observed in the literature, and also perhaps give more knowledge of the core ingredients needed to obtain such beats. 

While our discussion for studying non-reciprocal motion in active filaments has focussed thus far on biological contexts, a second application comes from biomimetic devices based on synthetic, filament-like colloidal structures that can be actuated by magnetic fields \citep{Elgeti2015PhysicsReview} or diffusiophoresis \citep{Michelin2023Self-PropulsionDroplets}. The follower force model is a simple forcing mechanism that can be used to provide insights into recovering self-sustained buckling dynamics. Based on values in the literature for phoretic particles \citep{Nishiguchi2018FlagellarField}, we can extract the particle radius and velocity to be $a = 3.17\mu $m and $V = 10\mu \text{ms}^{-1}$, respectively. Using the viscosity of water, $\eta = 1$cP, we can estimate the corresponding follower force by calculating the magnitude of the force required to keep a self-propelled colloid fixed, $F = 6\pi\eta a V \approx 5.98\times 10^{-13} ~\textrm{N}$. Using the bending stiffness reported for an artificial microswimmer in \citep{Dreyfus2005MicroscopicSwimmers}, $K_B \approx 3.3\times 10^{-22} \text{Jm}^{-1}$, we can extract the corresponding non-dimensional follower force for a chain of length $L = 50a$, i.e. a chain constructed of $25$ particles, to be $f \approx 45.5.$ This value is above the critical threshold, and so it is plausible that time-dependent dynamics can be induced in colloidal chains using available mechanisms. Furthermore, in the case where all the particles in the chain are active, the more relevant model would be a distribution of follower forces along the filament length which, following the same non-dimensionalisation as above, has a critical buckling value of $f \approx 4$ \citep{Ozdemir2025AModels,Clarke2025NonlinearFilaments}. In this case, we anticipate the chain would be able to access different states associated with higher forcing values. Thus, we suspect introducing asymmetries and anisotropies in the ways outlined in the paper would result in similar dynamics reported in this work.
%While this is too small to achieve buckling for a single follower force, increasing the chain length to $50$ particles, for example, leads to a follower force which would be sufficient to induce buckling of the colloidal chain. However in the case where all particles in the chain are active, the more relevant model would be a distribution of follower forces along the filament length which, following the same non-dimensionalisation as above, has a critical buckling value of $f \approx 4$ \citep{Clarke2025NonlinearFilaments}. Therefore it is plausible that the dynamics observed in these colloidal chains are a result of buckling. Thus, we suspect introducing asymmetries and anisotropies in the ways outlined in the paper would result in similar dynamics reported in this work. 

stiffnessThe anisotropies studied in this work are assumed to be constant along the filament length, but in reality this may not be the case. Towards the bottom of each cilium, connecting the axoneme to the basal body, is a region called the transition zone, in which several structural changes from the `9+2' axoneme can be found \citep{Mercey2024TheZones}. Incorporating the structural properties of this region is of interest, as defective proteins in the transition zone are known to cause a range of human diseases, called ciliopathies \citep{Goncalves2017TheGate}. Investigating the role of this coupling at the base, either through different boundary conditions at the wall or through spatially-varying anisotropies, will provide an insight to the impact of these localised anisotropies on ciliary dynamics. Another future avenue of interest is considering the collective motion of asymmetrically beating filaments, stiffnesssuch as in ctenophores, which use plates of synchronously beating cilia for locomotion \citep{Tamm2014CiliaCtenophores}. The plates of cilia coordinate such that a constant phase difference is observed between neighbouring plates, causing a metachronal wave along the ctenophore length. Each cilium has a `9+3' axoneme \citep{Afzelius1961TheSwimming-plates.}, suggesting anisotropy at the individual level is important to ctenophore motility. However it is not immediately clear how the structure of these comb plates, in particular how each cilium is bound to the plate, will affect both the individual and collective dynamics. The study of collective dynamics, in particular investigating if hydrodynamic interactions between cilia can lead to synchronisation and perhaps even metachronal waves as observed experimentally, will be the subject of our future research.

\section*{Declaration of Interests}
The authors report no conflict of interest.

\section*{Acknowledgement}
B.C. gratefully acknowledges funding from an EPSRC scholarship (Grant No. EP/W523872/1).

\appendix

\section{Initial buckling for $\kappa_0 \neq 0$}\label{appen:initialbuckling_alpha}
Recall that introducing asymmetry into the filament motion through intrinsic curvature separates the branches for the in-plane and off-plane buckling modes. Buckling off-plane occurs earlier for filaments with higher intrinsic curvature, while in-plane buckling occurs at the same forcing regardless of the curvature. It appears that this is associated to choice of frame in which the curvature is defined to act.

\begin{figure}
    \centering
    \includegraphics[width=\linewidth]{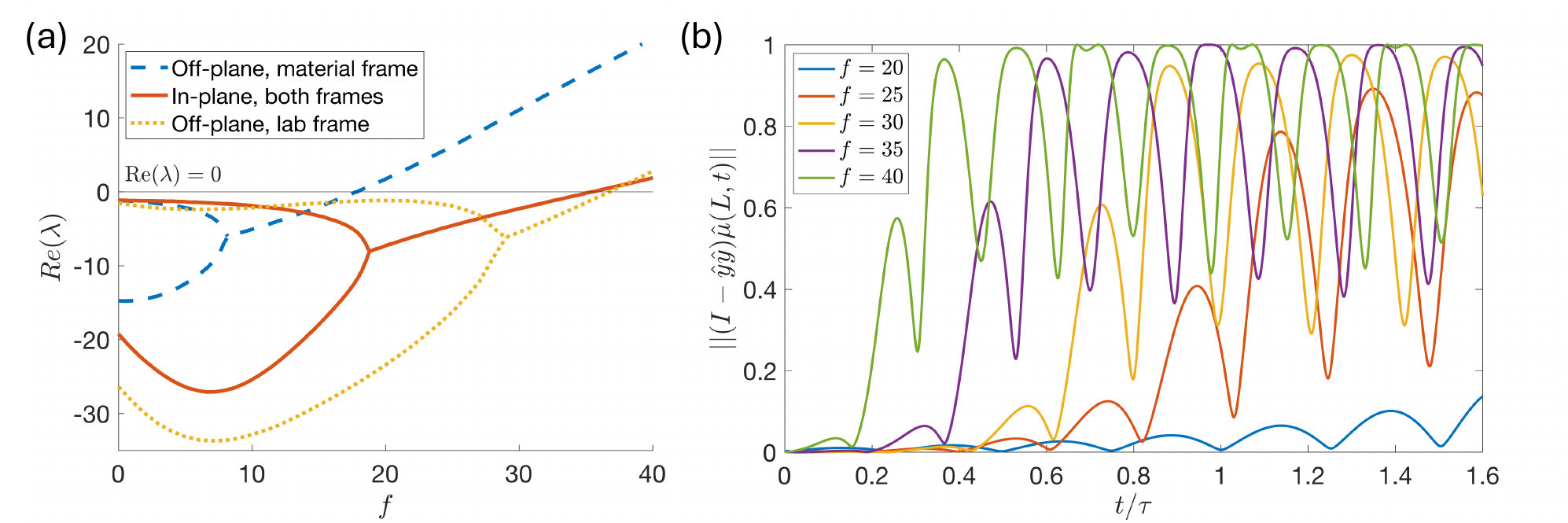}
    \caption{(a) The real part of the dominant four eigenvalues from a linear stability analysis of the steady state for $\kappa_0=3\pi/4,$ $ \beta=1$ associated to the off-plane buckling (blue dash/yellow dotted) or in-plane buckling (red) when the curvature is introduced in either frame. (b) The off-plane component of the direction of the intrinsic curvature for various values of the forcing. These are obtained from IVPs, as described in the text. We note that if the curvature is fixed in the lab frame, this would be zero for all time.}
    \label{fig:buckling_differentframes}
\end{figure}

We can choose the direction in which the curvature exists, i.e. in the material frame, so that the moment is given by
\begin{equation}\label{eq:moment_material}
    \bm{M}(s,t) = K_B \left(\hat{\bm{t}} \cdot \frac{\partial \hat{\bm{\nu}}}{\partial s} - L\kappa_0\right) \hat{\bm{\mu}} + \beta K_B\left( \hat{\bm{\mu}} \cdot \frac{\partial \hat{\bm{t}}}{\partial s}\right) \hat{\bm{\nu}}+ K_T \left( \hat{\bm{\nu}} \cdot \frac{\partial \hat{\bm{\mu}}}{\partial s}\right) \hat{\bm{t}},
\end{equation}
or in the lab frame, so that the moment is defined by 
\begin{equation}\label{eq:moment_lab}
    \bm{M}(s,t) = K_B \left(\hat{\bm{t}} \cdot \frac{\partial \hat{\bm{\nu}}}{\partial s}\right) \hat{\bm{\mu}} + \beta K_B\left( \hat{\bm{\mu}} \cdot \frac{\partial \hat{\bm{t}}}{\partial s}\right) \hat{\bm{\nu}}+ K_T \left( \hat{\bm{\nu}} \cdot \frac{\partial \hat{\bm{\mu}}}{\partial s}\right) \hat{\bm{t}} - K_B L\kappa_0 \hat{\bm{y}}.
\end{equation}
We show the stability analysis for both cases for $\kappa_0=3\pi/4$ in Figure \ref{fig:buckling_differentframes}a. Here, we see that in both cases, the in-plane buckling occurs at $f \approx 35.5$. When the intrinsic curvature is incorporated in the material frame, the off-plane buckling occurs much earlier at $f\approx17.8$, contrasting $f\approx 36.7$ for the lab frame case. This suggests that the earlier off-plane buckling arises due to the way in which the intrinsic curvature is tied to the filament frame; small perturbations to the filament generate changes to the filament's local frame, and thus to the direction of intrinsic curvature through $\hat{\bm{\mu}}.$ When the intrinsic curvature is defined in the material frame, a small off-plane change in $\hat{\bm{\mu}}$ leads to off-plane torque,
\begin{equation}
    \tau_\perp = - K_B\kappa_0 ||(\bm{I} - \hat{\bm{y}}\hat{\bm{y}}) \hat{\bm{\mu}}||.
\end{equation}
In essence, the filament now has a non-planar intrinsic curvature, causing off-plane torques which yield an earlier buckling. The effect of these torques is amplified as we increase $\kappa_0,$ explaining why the filament buckles earlier as we increase the intrinsic curvature.

In Figure \ref{fig:buckling_differentframes}b, we consider these off-plane effects by calculating the in-plane contributions of $\hat{\bm{\mu}},$ and hence of the intrinsic curvature. This data is taken from initial value problems in the case where the intrinsic curvature is defined in the material frame, i.e. using Eq \ref{eq:moment_material}. The initial filament configuration is given by the steady state, as found by JFNK, with an additional small perturbation of $\sim O(10^{-3}).$ We let this simulation run until periodic oscillations are achieved. As predicted by the stability analysis, the perturbations grow faster as $f$ increases, leading to  quicker saturation into the periodic beating we call P1. We can contrast this with the lab-frame intrinsic curvature, where the internal moment is defined by Eq \ref{eq:moment_lab}, in which the off-plane contribution of the torque is always $\tau_\perp = - K_B\kappa_0 ||(\bm{I} - \hat{\bm{y}}\hat{\bm{y}}) \hat{\bm{y}}|| = 0.$

%\section{Whirling and P1 for $\kappa_0 \in [0,3\pi/4], \beta=1$ }
%\begin{figure}
%    \centering
%    \includegraphics[width=\linewidth]{appendix_curvature_differentalpha.pdf}
%    \caption{Caption}
%    \label{fig:enter-label}
%\end{figure}

\section{Bifurcation Analysis}
In the following, we further discuss the bifurcation diagrams shown in the main text pertaining to isotropic filaments with intrinsic curvature, and anisotropic filaments with no intrinsic curvature, namely Figures \ref{fig:curvature_bifurcation_diagrams} and \ref{fig:bending_bifurcation_diagrams_and_stability}a.

\subsection{Isotropic filaments with intrinsic curvature}\label{appen:kappa0_bifurcations}
The bifurcation diagram for $\kappa_0 = 0$, i.e. the symmetric case, is shown in Figure \ref{fig:curvature_bifurcation_diagrams}a. As discussed in \citep{Clarke2024BifurcationsFilaments}, we see that the filament buckles at $f= f^*$, matching the planar case. At this point, both planar beating and whirling branches are born through a double Hopf bifurcation. Planar beating is initially unstable, and whirling initially stable. At $f\approx 137.3$, whirling becomes unstable through a supercritical Hopf bifurcation in which QP1 is born. At $f\approx 140.0$, a pitchfork bifurcation occurs, in which the QP1 branch terminates and planar beating becomes stable.

\begin{figure}
    \centering
    \includegraphics[width=\linewidth]{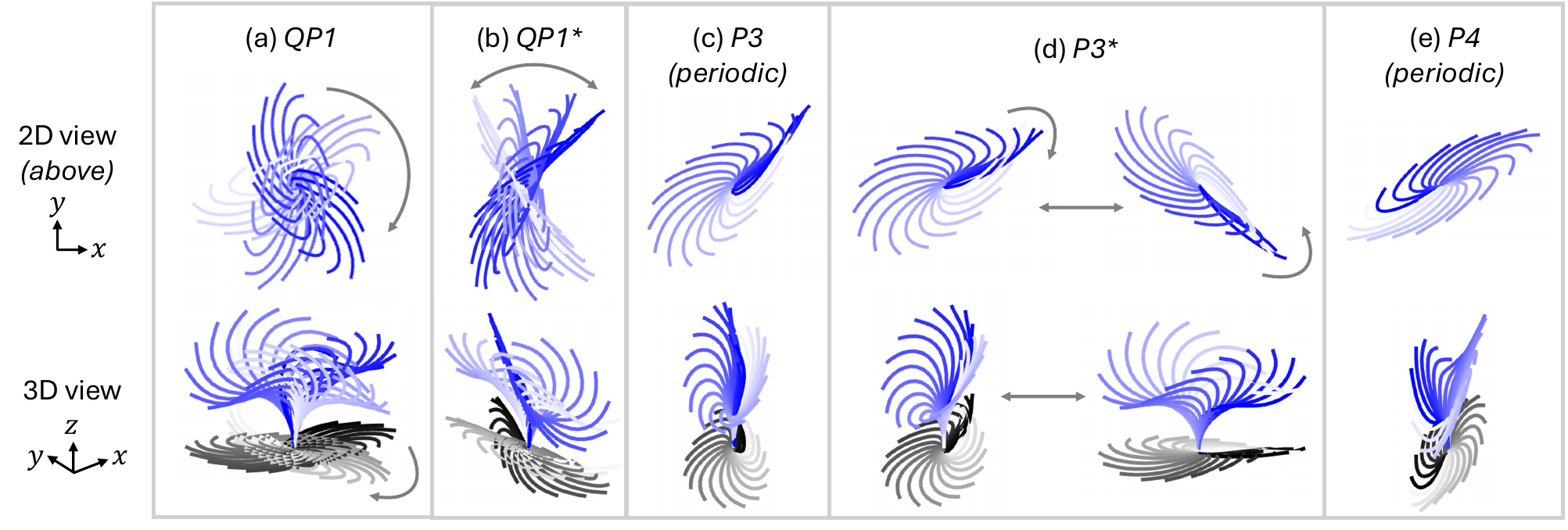}
    \caption{Plots of filament dynamics for solutions from two views. Darker lines correspond to moving forwards in time. Plots for (a) QP1, $(\kappa_0,\beta,f) = (0,1,138)$, (b) QP1*, $(\kappa_0,\beta,f) = (\pi/4,1,123)$, (c) P3, $(\kappa_0,\beta,f) = (\pi/2,1,80)$, (d) P3*, $(\kappa_0,\beta,f) = (\pi/2,1,90)$, (e) P4, $(\kappa_0,\beta,f) = (0,1.2,120)$. }
    \label{fig:appendix_extrasols}
\end{figure}

The bifurcation diagram for $\kappa_0=\pi/4$ is shown in Figure \ref{fig:curvature_bifurcation_diagrams}b. We see that the steady state becomes unstable earlier through a Hopf bifurcation at $f\approx 33.9 < f^*$, leading to a short region of $f$ in which only P1 is stable. We can also track the asymmetric planar beating branch found in the 2D case, which appears to be born at $f \approx f^*$ and remains unstable for all values of the forcing - agreeing with the IVPs. At $f\approx 36.3$, soon after the planar beating branch is born, the whirling solution branches from P1, exchanging stability through a pitchfork bifurcation (see Figure \ref{fig:curvature_bifurcation_diagrams}b, left inset). The whirling solution remains stable until $f\approx 126.4$. The P1 solution, however, becomes stable again at $f\approx 122.7$, leading to a region of bistability between the two. We perform bisection in this region, which unveils two distinct solutions (see Figure \ref{fig:curvature_bifurcation_diagrams}b, right inset); QP1, as in the isotropic case, and a new solution, which we call QP1*. We show these in Figure \ref{fig:appendix_extrasols}a,b. In this new solution, the filament has an almost-elliptical beat like QP1, although now the overall beat rotates from side-to-side, rather than unidirectionally like QP1. It is unclear whether there are bifurcations occurring between the unstable solutions in this region as bisection only allows us to compute the solution on the boundary between P1 and whirling.

For $\kappa_0 = \pi/2$, the bifurcation diagram is shown in Figure \ref{fig:curvature_bifurcation_diagrams}c. The steady states becomes unstable earlier at $f\approx 28.1$, giving rise to P1, and whirling branches off P1 later than the $\kappa_0=\pi/4$ case, at $f\approx 37.5$. Once more, we can find the asymmetric planar beating branch exists, and is unstable for all $f$ in this range. As in the previous case, we see a region of bistability between whirling and P1, but now for a larger range of $f$ (for $73.3\lesssim f \lesssim 100.3$). We again use bisection in this region, which reveals three solution branches (see Figure \ref{fig:curvature_bifurcation_diagrams}c, right inset). The first branch is a periodic solution close to the P1 bifurcation which we call P3; here the tip traces a tear-drop shape in the $(x,y)-$plane, as shown in Figure \ref{fig:appendix_extrasols}c. For larger $f$, bisection instead reveals the second branch, separated by the black line in Figure \ref{fig:curvature_bifurcation_diagrams}c inset; a solution which we call P3*. Here, the filament beats like P3, but now the tear-drop shape traced by the filament tip moves with time. The beat repeatedly changes direction, each time beginning to align the major axis of the shape traced by the filament tip with the $\hat{\bm{x}}-$axis before changing direction again, as illustrated in Figure \ref{fig:appendix_extrasols}d. The final branch, close to the whirling bifurcation, is QP1 as in the isotropic case. We note that the bifurcations corresponding to the loss and subsequent gain of stability of P1 are closer than the $\kappa_0=0$ and $\kappa_0 = \pi/4$ cases.

By $\kappa_0 = 3\pi/4$, these two pitchfork bifurcations have coalesced and vanished, as can be seen by the bifurcation diagram in Figure \ref{fig:curvature_bifurcation_diagrams}d. Here, the steady state becomes unstable at $f \approx 17.8$. P1 then remains stable beyond $f=300$, and the planar beating branch is again unstable for all $f$ in this range. Notably, the whirling solution is not only unreachable through IVPs for $\kappa_0=3\pi/4$, but appears to have vanished completely. A discussion on how and why this disappearance occurs can be found in the main text.

\subsection{Anisotropic filaments with zero-curvature rest state}\label{appen:beta_bifurcations}
The bifurcation diagram for $\beta = 1.2$ is shown in Figure \ref{fig:bending_bifurcation_diagrams_and_stability}a (top). We find that the steady state becomes unstable at $f = f^*,$ analogous to the $\beta = 1$ case. At this point, the in-plane beating branch (corresponding to beating in the $(x,z)-$plane) is born through a Hopf bifurcation. The off-plane beating branch (corresponding to beating in the $(y,z)-$plane) bifurcates from the trivial steady state at $f\approx \beta f^* = 53.0$, and is unstable for all values of $f.$ In-plane beating becomes unstable at $f\approx 46.0$, and the whirling branch is born. The whirling branch is then stable until $f \approx 138.4$, where it becomes unstable through a subcritical Hopf bifurcation.  At $f\approx 106.2$ the in-plane beating branch becomes stable, leading to a region of bistability between whirling and in-plane beating. Bisection in this region yields a single periodic solution branch; P4, in which the filament undergoes elliptical whirling oscillations, where the major axis of the ellipse has components in both the  $\hat{\bm{x}}$ and $\hat{\bm{y}}$ planes. We plot P4 over one period in Figure \ref{fig:appendix_extrasols}e. Tracking P4 with JFNK shows that P4 connects to the off-plane beating branch at $f \approx 178$. It is likely that other unstable solutions are present in this region of bistability, in particular connecting from the bifurcation in which whirling becomes unstable, which we are unable to obtain with bisection alone.

The bifurcation diagram for $\beta = 1.3$ is shown in Figure \ref{fig:bending_bifurcation_diagrams_and_stability}a (middle). There are many qualitative similarities to the previous bifurcation diagram; in-plane beating and off-plane beating both bifurcate from the trivial state at $f= f^*$ and $f=\beta f^*$ respectively, off-plane beating is always unstable, and in-plane beating becomes unstable as whirling becomes stable at $f\approx54.0$. We again observe a region of bistability between whirling and in-plane beating, now for $93.9 \lesssim f \lesssim 110.8$, in which we see the same unstable periodic solution, P4. However, a key difference is observed; P4 now connects to the whirling branch, after which both branches terminate.

The bifurcation diagram for $\beta = 1.4$ is shown in Figure \ref{fig:bending_bifurcation_diagrams_and_stability}a (bottom). By this value of $\beta,$ the bifurcations corresponding to in-plane beating becoming unstable, restabilising and the whirling branch have collided, and we see only one stable branch - in-plane beating. This bifurcates from the steady state at $f = f^*.$ The off-plane beating bifurcates from the steady state at $f = \beta f^*$, and is unstable for all values of the forcing.
\begin{figure}[t]
    \centering
    \includegraphics[width=0.8\linewidth]{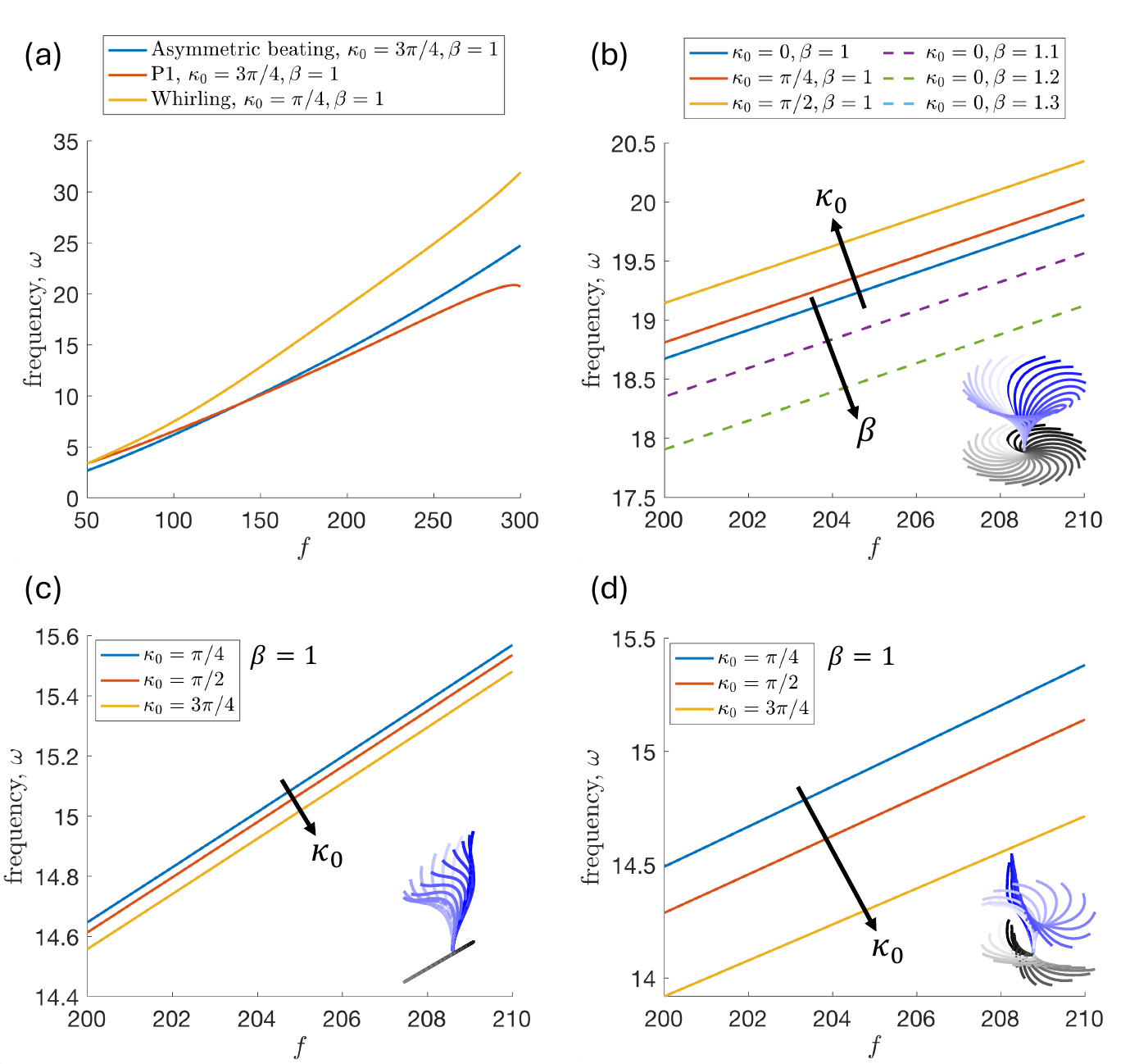}
    \caption{(a) The frequencies, $\omega = 1/T$ where $T$ is the solution period obtained via JFNK, of three solution branches for $f\in[50,300]$. We show asymmetric beating for $(\kappa_0,\beta) = (3\pi/4,1),$ P1 for $(\kappa_0,\beta) = (3\pi/4,1),$ and whirling for $(\kappa_0,\beta) = (\pi/4,1).$ The frequencies for various $\kappa_0$ and $\beta$ values for (b) whirling, (c) asymmetric beating and (d) P1 solutions are also shown for $f\in [200,210],$ to show the small variations in frequency as these parameters change.}
    \label{fig:frequencies}
\end{figure}

\section{Solution Frequencies}\label{appen:frequencies}

In Figure \ref{fig:frequencies}a, we plot the frequencies, $\omega = 1/T$, of some of the key dynamics observed in the main text; asymmetric beating, P1, and whirling. These are each plotted for a single choice of $\kappa_0$ and $\beta$ for $f\in [50,300]$. While the frequency of the solutions vary considerably with the follower force, as shown in Figure \ref{fig:frequencies}a, the change with the other parameters is much smaller, as shown in Figures \ref{fig:frequencies}b-d. The frequency of the whirling solutions increases with $\kappa_0$ and decreases with $\beta$ (Figure \ref{fig:frequencies}b) while the frequency of the asymmetric beating and P1 solutions both decrease with $\kappa_0$ (Figures \ref{fig:frequencies}c and d, respectively).

\section{Supplementary videos}
Supplemental videos are provided to highlight some of the key dynamics discussed in the text.
\begin{enumerate}
    \item \textit{sm1\_tilted\_whirl.mp4} Tilted whirling for $(\kappa_0,\beta,f) = (\pi/2,1,45).$
    \item \textit{sm2\_p1.mp4}: P1 for $(\kappa_0,\beta,f) = (3\pi/4,1,200).$
    \item \textit{sm3\_elliptical\_whirl.mp4}: Elliptical whirling for $(\kappa_0,\beta,f) = (0,1.2,46).$
    \item \textit{sm4\_asymmetric\_beating.mp4}: Asymmetric beating for $(\kappa_0,\beta,f) = (3\pi/4,2,40).$
\end{enumerate}

\bibliographystyle{unsrt}
\bibliography{references}

\end{document}